\newif\ifsubmode
\submodefalse

\newif\ifprintfig
\printfigtrue

\newif\ifemulate
\emulatetrue

\ifsubmode
      \documentclass[12pt,preprint]{aastex}
      \usepackage{psfig}
      \slugcomment{Submitted to Ap.J.}
      \shortauthors{Dalcanton et al.}
      \shorttitle{The Formation of Dust Lanes}
      \slugcomment{Submitted to Ap. J.}
\else
      \ifemulate
	     \documentclass{article}
	     \usepackage{emulateapjAASTeX4}
	     \usepackage{onecolfloat}
	     \usepackage{graphicx} 
	     \lefthead{Dalcanton et al.}
	     \righthead{The Formation of Dust Lanes}
	     \submitted{{\it Accepted for publication in ApJ}}
      \else
	     \documentstyle[11pt,psfig,aaspp4]{article}
	     \tightenlines
      \fi
\fi

\newcommand\hi{\noindent \hangindent=2.5em}
\newcommand\degree{{^\circ}}
\newcommand\pc{{\rm\,pc}}

\newcommand\kpc{{\rm\,kpc}}

\newcommand\kmsec{{\rm\,km\,s^{-1}}}
\newcommand\kms{\kmsec}

\newcommand\K{{\rm\,K}}

\newcommand\msun{{\rm\,M_\odot}}
\newcommand\lsun{{\rm\,L_\odot}}

\newcommand\clock{\count0=\time \divide\count0 by 60
     \count1=\count0 \multiply\count1 by -60 \advance\count1 by \time
     \number\count0:\ifnum\count1<10{0\number\count1}\else\number\count1\fi}

\begin{document}

\ifemulate
   \twocolumn[
\fi

\title{The Formation of Dust Lanes: Implications for Galaxy Evolution}

\author{Julianne J. Dalcanton\altaffilmark{1,2}, Peter Yoachim\altaffilmark{3}}
\affil{Department of Astronomy, University of Washington, Box 351580,
Seattle WA, 98195}

\author{Rebecca A. Bernstein\altaffilmark{4}}
\affil{Department of Astronomy, University of Michigan, Ann Arbor MI, 48109}

\altaffiltext{1}{e-mail address: jd@astro.washington.edu}
\altaffiltext{2}{Alfred P.\ Sloan Foundation Fellow}
\altaffiltext{3}{e-mail address: yoachim@astro.washington.edu}
\altaffiltext{4}{e-mail address: rabernst@umich.edu}
  
\begin{abstract}

{\small{ From a survey of edge-on disks, we find that disk galaxies
show a sharp, mass-dependent transition in the structure of their
dusty ISM.  In more massive, rapidly rotating disks with
$V_c>120\kms$, we see the well-defined dust lanes traditionally
associated with edge-on galaxies.  However, in more slowly rotating,
lower mass galaxies with $V_c<120\kms$, we find {\emph{no}} dust
lanes.  Instead, the distribution of dust in these galaxies has a much
larger scale height and thus appears more diffuse.  Evidence suggests
that the change in scale height is due primarily to changes in the
turbulent velocities supporting the gas layer, rather than to sharp
changes in the gas surface density.  A detailed analysis of our sample
reveals that the decrease in the dust scale height is associated with
the onset of disk instabilities, evaluated for a mixed star$+$gas
disk.  Specifically we find that all of the high-mass galaxies with
dust lanes are gravitationally unstable, and thus are prone to
fragmentation and gravitational collapse along spiral arms.
Empirically, our data implies that turbulence has lower characteristic
velocities in the presence of disk instabilities, leading to smaller
gas scale heights and the appearance of narrow dust lanes.  The drop
in velocity dispersion may be due either to a switch in the driving
mechanism for turbulence from supernovae to gravitational
instabilities or to a change in the response of the ISM to supernovae
after the ISM has collapsed to a dense layer.  We hypothesize that the
drop in gas scale height may lead to significant increases in the star
formation rate when disk instabilities are present.  First, the
collapse of the gas layer increases the typical gas density, reducing
the star formation timescale.  Second, the star formation efficiency
increases due to lower turbulent velocities.  These two effects can
combine to produce a sharp increase in the star formation rate with
little change in the gas surface density, and may therefore provide an
explanation for the Kennicutt surface density threshold for star
formation.  Our data also suggest that star formation will be
systematically less efficient in low mass disks with $V_c<120\kms$,
since these galaxies are stable and lie entirely below the Kennicutt
surface density threshold.  In these stable systems the effective
nucleosynthetic yield is reduced because the star formation timescale
becomes longer than the gas accretion timescale, supressing the
metallicity.  This effect can possibly produce the observed fall-off
in metallicity at rotation speeds less than $V_c<120\kms$.  Thus,
infall provides an equally plausible explanation of the
mass-metallicity relation in disks as global outflows driven by supernova
winds.  The transitions in disk stability, dust structure, and/or star
formation efficiency may also be responsible for the observed changes
in the slope of the Tully-Fisher relation, in the sharp increase in
the thickness of low mass galaxy disks, and in the onset of bulges in
galaxies with $V_c\gtrsim120\kms$.  The latter observation lends
support to theories in which bulges in late-type galaxies grow through
secular evolution in response to disk instabilities.  We include in
this paper relationships between the surface density and the vertical
stellar velocity dispersion as a function of galaxy rotation speed,
which may be useful constraints on galaxy formation models.  }}

\end{abstract}

\ifemulate
    ]
    \setcounter{footnote}{0}
\fi

\section{Introduction}

Dust lanes have long been recognized as a characteristic signature of
edge-on disk galaxies.  Optical images show that edge-on disks are
typically bisected by a dark narrow layer of dust which has a typical
scale height half as large as that of the stellar disk (Xilouris et
al.\ 1999).  However, not all galaxies show this conspicuous feature.
In many late-type, low mass galaxies the dust lane is absent.  It has
been assumed that the lack of a dust lane reflects an overall lack of
dust, possibly due to low metallicity (van den Bergh \& Pierce 1990).
However, low mass galaxies {\emph{do}} have dust (see compilation in
Lisenfeld \& Ferrara 1998).  The difference lies in the morphology of
the dust distribution, which is less concentrated within the disk and
significantly more porous (see e.g.\ Matthews \& Woods 2001).

We are engaged in an extensive multicolor imaging program of edge-on
bulgeless disks (Dalcanton \& Bernstein 2000) that is ideal for
probing variations in dust lane morphology.  Our study of this sample
indicates that the transition from a thicker, more porous distribution
of dust to a narrow lane is remarkably sharp.  The dramatic change in
the morphology of the dusty ISM with increasing galaxy mass cannot be
due solely to increasing metallicity or dust content, but rather must
result from the onset of a distinct physical mechanism.

As discussed in more detail in \S\ref{theorysec}, dust is an excellent
tracer of gas in galaxies, with the regions of highest extinction
tracing regions of the highest gas density, namely the cool neutral
and the cold molecular medium.  This correlation is a clue to the
observed transition in the dust distribution, and suggests that the
dust morphologies indicate a change in the structure of
the cool and cold ISM.  This connection allows us to use the
transition in dust morphology to diagnose the processes which shape
the global distribution of the cold ISM.

The outline of the paper is as follows.  In \S\ref{datasec} we present
the evidence in our sample for a sharp change in the dust lane
morphology with decreasing mass.  We suggest that the observed
disappearance of dust lanes results from a sharp increase in the
thickness of the cold ISM in low mass galaxies.  In \S\ref{quantsec}
we discuss the physics which might drive this morphological dichotomy,
and find that disk stability is the only intrinsic property that
changes sharply with dust morphology.  In \S\ref{theorysec}, we discuss the
relationship between gravitational instabilities and interstellar
turbulence, and describe how it might lead to a thinner layer of cold
ISM in unstable high mass galaxies.  In \S\ref{SFsec} we discuss the
implications of our observations for the characteristic scales and
velocities of turbulence driven by gravitational instabilities, and
suggest that these may provide an explanation for the Kennicutt star
formation threshold.  In \S\ref{metallicitysec}, we discuss how the
resulting drop in star formation efficiency in low mass disk galaxies leads
to lower metallicities and nucleosynthetic yields at the same
characteristic velocity where dust lanes disappear, without resorting
to global galactic outflows. In \S\ref{observationsec} we discuss how
connections between dust morphology, disk stability, and star
formation efficiency may be evident in other observations, such as the
Tully-Fisher relation, the bulge-disk ratio as a function of galaxy
mass, the thickness of dwarf galaxies, and the properties of galaxies
at high redshift.  We summarize our conclusions in
\S\ref{conclusionsec}.

\section{The Mass Dependence of Dust Lane Morphology} \label{datasec}

In two previous papers (Dalcanton \& Bernstein 2000a, hereafter
Paper I, and Dalcanton \& Bernstein 2002, hereafter Paper II) we
describe the sample selection, image reduction, and extraction of
vertical color gradients from a sample of 49 edge-on
bulgeless disk galaxies spanning a wide range in rotation speed,
or equivalently, mass.  The $B$, $R$, and $K_s$ images are
presented in Paper I.  The $B-R$ and $R-K_s$ color maps and
vertical color gradients are discussed Paper II.

\begin{figure}[t]
\includegraphics[width=3.5in]{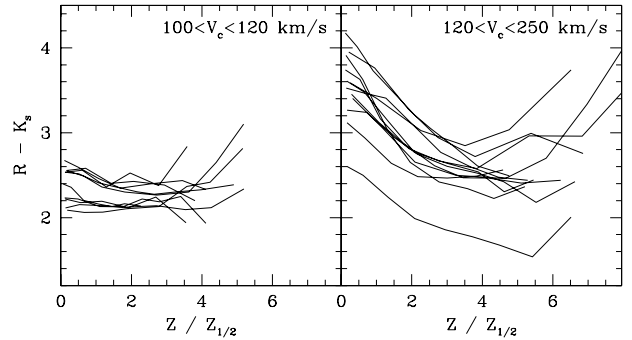}
\caption{\footnotesize 
  The dust-sensitive $R-K_s$ color as a function of height above
  the midplane, in two bins of galaxy rotation speed $V_c$.  All
  galaxies with $V_c>120\kms$ (right panel) have strong color
  gradients and become dramatically bluer above the midplane,
  indicating the presence of a confined dust lane.  In contrast, all
  the low mass galaxies ($V_c<120\kms$) show no color gradients.  The
  distance above the midplane is scaled by the disks' half-light scale
  height $z_{1/2}$, measured in the $K_s$-band.
  \label{dustfig}}
\end{figure}

While analyzing the data for Paper II, we noticed a remarkable fact.
As is visually apparent in the dust sensitive $R-K_s$ maps shown in
Paper II, all galaxies with high rotation speeds ($V_c>120\kms$) have
obvious dust lanes, while {\emph{none}} of the galaxies with lower
rotation speeds ($V_c<120\kms$) do.  We have quantified this change in
dust morphology by plotting the vertical color profiles as a function
of the galaxies' rotation speed $V_c$, as shown in
Figure~\ref{dustfig}.  Here, $V_c$ is derived from either the galaxies' HI
line widths or from their optical rotation curves when HI data was not
available.  The colors were measured using the average flux within
$\pm$1 radial exponential scale length of the center of the galaxy,
averaging above and below the plane, as discussed in Paper II.  We
have excluded two galaxies (FGC E1619 and FGC 2217) because they are
not exactly edge-on, and therefore have significantly different color
gradients above and below the plane.  Errors were calculated in a
Monte Carlo fashion to account for the Poisson errors due to flux
uncertainties and the correlated errors due to sky subtraction,
following Bell \& de Jong (2000).  Data for which the uncertainty in
$R-K_s$ is above 0.5$^m$ are not shown.  Such large errors in color
typically occur only at the furthest point from the midplane (see
Figure 3 of Paper II).  The distances above the plane were scaled by
the galaxies' $K_s$-band half-light height $z_{1/2}$, defined so that half of a
galaxies' light is contained between $\pm z_{1/2}$.

\placefigure{dustfig}

The right hand panel of Figure~\ref{dustfig} shows the vertical color
profiles of all galaxies with $V_c>120\kms$.  These galaxies all show
steep declines in $R-K_s$ color ($>1^m$) with increasing scale
heights, due the presence of a well-defined dust lane.  The regions
close to the midplane are highly obscured at optical wavelengths,
leading to colors that are far redder than any realistic stellar
population.  Outside of the dust lane, the reddening is much lower,
revealing the bluer colors of the underlying stellar population and
suggesting that the dust lane is well-confined within the stellar
disk, as found in other detailed modeling studies (Xilouris et al.\
1999, and references therein).  All of the variations in the vertical
color profiles are similar in amplitude and extent, suggesting a large
degree of uniformity in the structure of the dust lanes\footnote{The
one apparent exception, FGC 436, is at very low galactic latitude.
Given the similarity in shape between its color profile and the
others, it seems likely that the true Galactic extinction toward this
galaxy is less than we adopted in Paper II.  The $B-R$ colors of the
galaxy are also consistent with this interpretation.}.

In contrast, the vertical color profiles of more slowly rotating lower
mass galaxies ($100<V_c<120\kms$; left panel) are distinctly
different.  In these, we can detect no color gradients within
$\pm0.2^m$, consistent with the visual impression that these galaxies
have no dust lanes.  In a few galaxies there is a hint of a slight
gradient ($\Delta(R-K_s)<0.2^m$), but it is neither statistically
significant nor reflected in the $B-R$ color gradients (see Figures 3
and 5 of Paper II).  Galaxies of even lower mass ($V_c<100\kms$)
likewise show no evidence for dust lanes.  They have color gradients
comparable to those shown here, or become increasingly {\emph{redder}}
above the plane.

Figure~\ref{dustfig} strongly suggests that $V_c=120\kms$ represents a
dividing line for the dust properties of disk galaxies.  There is
absolutely no overlap in the properties of the $R-K_s$ color profiles
between the two regimes, despite the small difference in rotation
speed.  (There are less than $10\kms$ separating the maximum rotation
speed in the left panel and the minimum in the right.)  We would
expect that this transition might be somewhat less sharp within a more
heterogeneous sample than ours.  However, we have inspected images
from the Sloan Digital Sky Survey Early Data Release (Stoughton et al
2002) of another $\sim30$ edge-on galaxies with known circular
velocities and found identical results.  This suggests that there are
two distinct regimes for the morphology of the dust in disk galaxies
and that the transition between these regimes is a very sharp function
of galaxy mass.

While there are alternate possibilities explaining the trend seen in
Figure~\ref{dustfig}, none of these seem likely when combined with
other observations.  One alternate possibility is that low mass
galaxies do have dust lanes, but they are obscured by current star
formation near the midplane.  This would also cause a strong increase
in the $B-R$ color toward the midplane due the young population.
However, the $B-R$ color gradients of the low mass systems are as flat
as those in $R-K_s$, as shown in Paper II.  The simplest explanation
is that the stellar populations and the extinction are quite uniform
over a large range in height, not that there is a population of stars
which just happens to exactly fill in the dust lane.

\begin{figure}[t]
\includegraphics[width=3.5in]{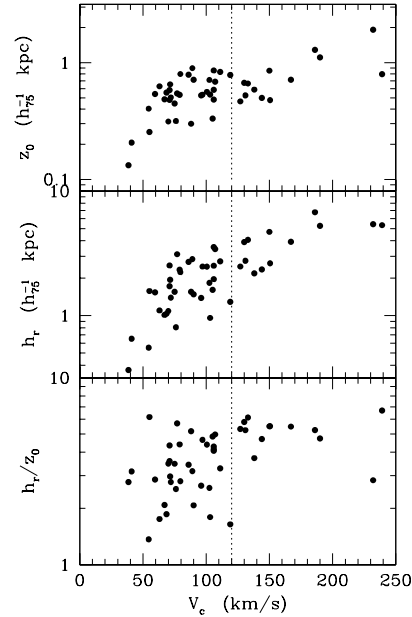}
\caption{\footnotesize 
  $K_s$-band structural parameters derived from 2-dimensional
  fits of edge-on exponential disks, as a function of rotation speed
  for the galaxies in the Dalcanton \& Bernstein (2000) sample.  The
  top panel shows the vertical scale height $z_0$ of the disks, and
  demonstrates that there is substantial overlap in the vertical extent
  of the stellar disks above and below the $V_c=120\kms$ limit.  The middle
  panel shows the exponential scale lengths $h_r$ of the disks.  The
  bottom panel shows the axial ratio of the disks ($h_r/z_0$).  The
  disks with $V_c<120\kms$ have systematically smaller axial ratios.
  \label{structparamfig}}
\end{figure}

A second alternate explanation of the data in Figure~\ref{dustfig} is
that the low mass galaxies host dust lanes of comparable thickness to
those in the high mass galaxies, but that the stellar disk is
sufficiently thin that it is entirely embedded within the dust layer and
thus no $z$-gradient appears.  This explanation fails for several
reasons.  First, the stellar disk must be substantially smaller than
the dust lane to produce a color gradient as small as we observe.  If
we admit a generous color gradient of $\Delta(R-K_s)<0.3^m$ for the
low mass galaxies, then these must have a stellar scale height,
$z_{1/2}$, that is more than a factor of three times smaller than the
high mass galaxies (assuming that the strength of the dust lane is
similar), in spite of the fact that the mass of the galaxies has
changed very little.  Second, we can do the more direct test of
measuring the actual stellar scale heights for the galaxies.  Based on
2-dimensional fits of exponential disks viewed in projection (Yoachim
\& Dalcanton 2003), we find that there is in fact substantial overlap
in the $K_s$-band scale heights of the low and high mass subsamples,
as shown in the top panel of Figure~\ref{structparamfig}.  This was
initially somewhat surprising, as we expected the lower mass galaxies
to indeed be smaller.  However, as the bottom panel of
Figure~\ref{structparamfig} shows, $V_c=120\kms$ also seems to be the
point at which the stellar disks become proportionally
{\emph{thicker}} compared to their radial scale lengths, which
compensates for the expected decrease in the size of the galaxy.  (We
will return to this point in \S\ref{thicksec}, as we believe it is
closely related to the disappearance of the dust lane.)  As a final
test that changes in the scale height of the stellar disk are not
responsible for masking the true presence of the dust lane, we inspected
plots of the color profiles as a function of distance above the plane,
rather than scaled to the disks' scale height.  The color profiles of
the low mass galaxies can easily be traced to large enough distances
that the steep color gradient seen in high mass galaxies would be
readily visible, if present.

\placefigure{structparamfig}

While the evidence suggests that galaxies with $V_c<120\kms$ do not
have dust lanes, it does not necessarily suggest that there is a large
drop in the overall quantity of dust in these galaxies. Dust is
probably a by-product of late-stage stellar evolution (Fleisher,
Gauger, \& Sedlmayr 1992, Gail \& Sedlmayr 1988), and thus even low mass
galaxies should also have produced dust.  Previous studies of
extinction and dust-to-gas ratios do suggest that this is the case, as
they have found a steady decrease in reddening with decreasing galaxy
mass and/or luminosity (e.g.\ Stasinska \& Sodr\'e 2001, Zaritsky et
al.\ 1994), but not a sharp drop in the inferred dust-to-gas ratio.

It follows that dust is still present in galaxies with $V_c<120\kms$,
but that the spatial distribution of dust has altered dramatically.
Specifically, the data are consistent with the dust having a much
larger scale height and thus more porous spatial distribution and
lower line-of-sight opacity when viewed edge-on.  The ground-based
images of our sample (Paper I) frequently show significant patchy dust
extinction at all scale heights in the low mass galaxies.  This
confirms that dust is present but is no longer confined to a thin
layer.  Although almost no late-type edge-on galaxies have published HST
observations, the one that has been fully analyzed also shows a
significant dust component with a clumpy distribution.  A detailed HST
study of one slowly rotating, edge-on, late-type galaxy comparable to
those in our sample (UGC 7321 with $V_c=109\kms$; Matthews \& Wood
2001) reveals a non-negligible amount of dust distributed in clumps
and filaments throughout the galaxy, rather than confined to a thin
dust lane.

\begin{figure*}[t]
\hbox{ 
\includegraphics[width=3.5in]{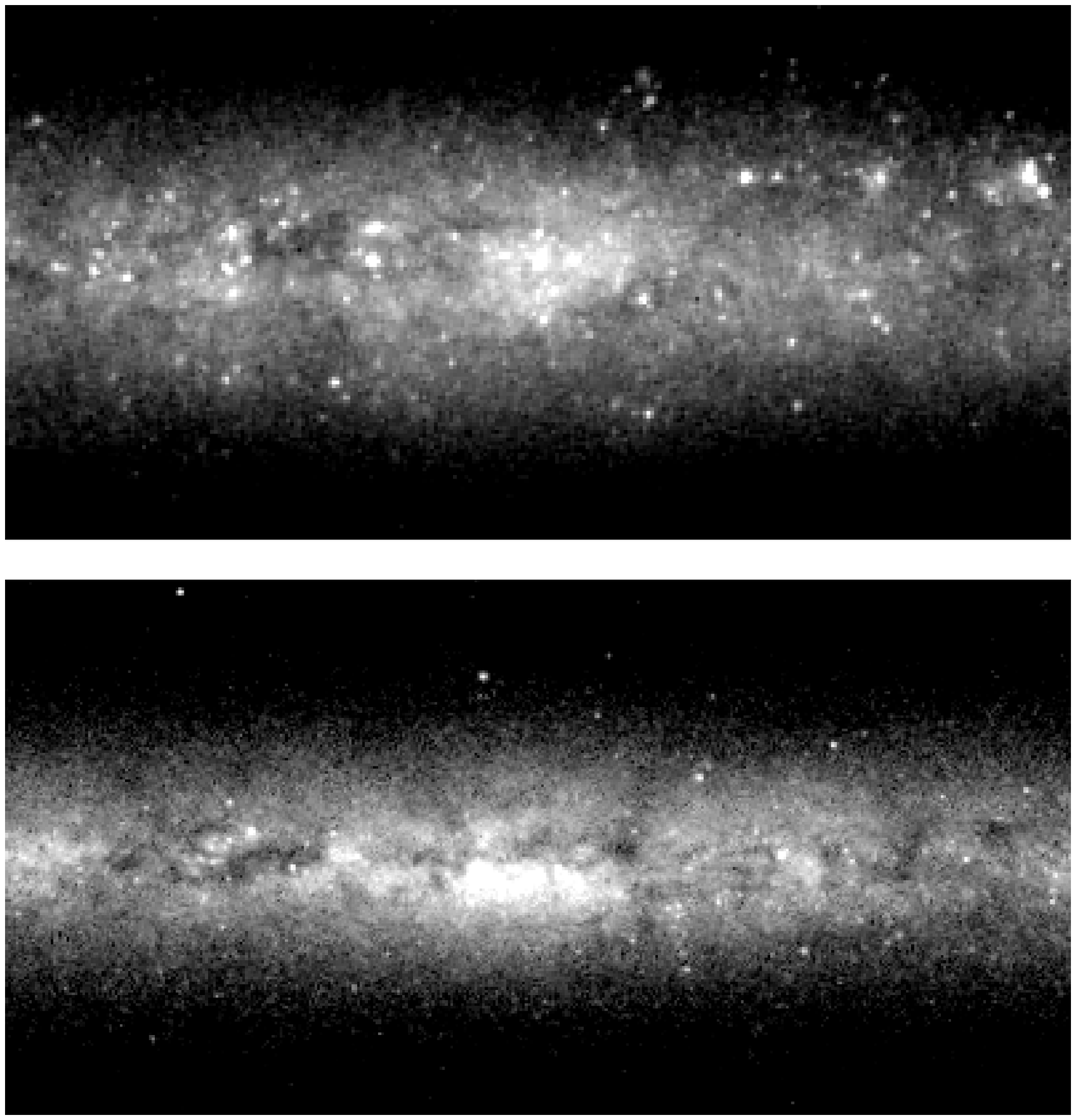}
\includegraphics[width=3.5in]{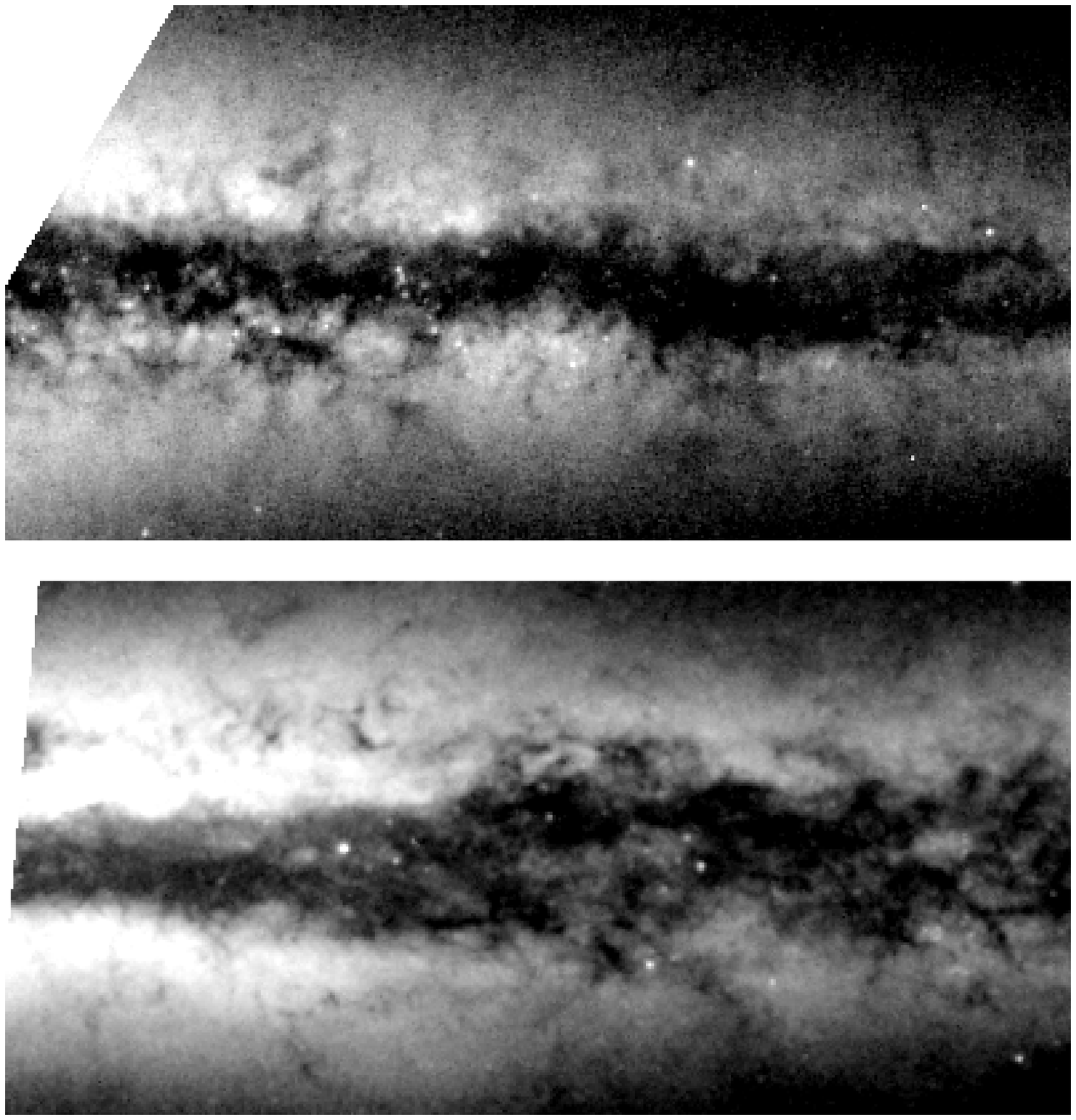}}
\caption{[a; left] HST WFC images of two low mass late-type galaxies
  (UGC 711, $V_c=101\kms$, $F702W$ filter [top] and UGC 7321,
  $V_c=105\kms$, $F814W$ filter [bottom]).  In neither galaxy is the
  dust particularly concentrated near the midplane.  Instead, large
  amounts of dust can be detected at large scale heights, out to the
  surface brightness limit of the data.  The images have been scaled
  to the same distance and are approximately centered on the galaxy.
  The horizontal extent of the image is $\sim3\kpc$, and thus the
  images extend to $\sim1.5\kpc$ in radius from the center of the
  galaxies, or $\lesssim1$ disk scale length.  [b; right] HST WFC
  images of two high mass late-type galaxies (NGC 4013, $V_c=196\kms$,
  $F814W$ filter [top] and NGC 4302, $V_c=180\kms$, $F814W$ filter
  [bottom]).  The images have been scaled to the same distance as the
  galaxies in Figure~\ref{hstfig}[a].  However, due to the location of
  the galaxies on the WFC chips, these two images are {\emph{not}}
  centered on the galaxies.  Instead, the center of the galaxies are
  approximately on the left hand edge of the image, and thus, the
  images extend to larger radii than the images of the low mass
  galaxies ($r\sim 3\kpc$ vs $r\sim1.5\kpc$), but to a similar number
  of disk scale lengths.  The centers of both galaxies have narrow
  well-defined dust lanes, in sharp contrast to the low mass galaxies.
  However, outside the inner region of NGC 4302, the distribution
  becomes much thicker and more diffuse, with the transition in
  morphology taking place over a narrow range in radius.  The
  distribution of dust in this outer region may be comparable to the
  distribution of dust in the low mass galaxies.  
  \label{hstfig}}
\end{figure*}

To verify that a change in the dust scale height is responsible for
the disappearance of dust lanes we have extracted images of several
other galaxies from the HST Archive ``Associations''.  We have chosen
these galaxies to be very late-type and completely edge-on, comparable
to those in our sample.  There are only two suitable observations of
galaxies with $V_c<120\kms$ (due to few studied galaxies, low
signal-to-noise, and/or galaxies' being resolved into stars), and not
many more with $V_c>120\kms$ (due to the presence of large bulges, and
significant deviations from $i=90\degree$).  In Figure~\ref{hstfig} we
show images of two galaxies from each mass range, in the reddest
filter available.  We have scaled the sizes of all four images to the
same effective distance using distances from Tully (1988) and from
Karachentsev et al.\ (2000; UGC 7321) so that the true thicknesses of
the dust lanes can be directly compared.  Figures~\ref{hstfig}a\&b
show the two low mass and high mass galaxies, respectively.  Due to
limitations of the archival data, the images of the low mass galaxies
are centered on the galaxies, while the high mass galaxy images are
shifted radially so that the center of the galaxy falls on the left
edge of the image.  Therefore, while all of the HST images show a
$\sim3\kpc$ region of the disk, the images of the high mass galaxies
extend twice as far in radius.

\placefigure{hstfig}

The HST imaging reveals clear differences in the distribution of dust
between low and high mass galaxies.  The central dust lanes of the
high mass galaxies are extremely compact vertically, with relatively
sharp boundaries above and below the lane.  While these galaxies do
have some dust absorption above the plane (and indeed, both were
selected for HST study on that basis), the extraplanar dust is clearly
a very small fraction of the total.  In contrast, the low mass
galaxies do not show evidence for an equally compact lane, even one
with significantly lower opacity.  Although the imaging for the low
mass galaxies is not as deep, the observations are sufficient to
eliminate the possible existence of a compact dust lane.  Note that
they do, however, have significant dust absorption, albeit patchier
and with a larger scale height\footnote{The radiative transfer
modeling of UGC 7321 by Matthews \& Wood (2001) did not attempt to
model the scale height of the dust, and instead adopted a height of
1/2 that of the stars, as found by Xilouris et al.\ (1999) for
galaxies with well-defined dust lanes.}.  The HST data and our own
observations therefore suggest that the lack of dust lanes in low mass
galaxies is best explained by an increase in the scale height of the
dust, rather than a sharp decrease in the relative dust mass.

Although the dust morphologies in the centers of the galaxies are
distinct, there is one morphological similarity between the
distributions of dust in Figures~\ref{hstfig}.  Namely, in the
{\emph{outer}} regions of the high mass galaxy NGC 4302 (lower panel),
the distribution of dust is far more porous and has a noticeably
larger scale height than in the center of the galaxy.  The dust in the
outer regions is distributed with a morphology similar to those in the
low mass galaxies, but perhaps with a higher overall opacity.  In NGC
4302, the transition in the distribution of dust occurs rapidly over a
very small range radius, possibly reflecting a change in physical
conditions {\emph{within}} the disk comparable to the change from
galaxy-to-galaxy at $V_c=120\kms$.

\section{ISM Structure and Disk Instability}  \label{quantsec}

Rotation speed (i.e.\ mass) is unlikely to be the parameter which
directly controls the physics of the transition in the dust scale
height.  Instead, the transition is likely to originate from one or
more of the correlations found between the properties of disk galaxies
and their rotation speeds.  We find that several of the parameters
that have previously been suggested as crucial for driving variations
in the ISM, such as luminosity density, have relatively weak or
non-existent trends with rotation speed within our
sample\footnote{Note, however, that we considered only mean properties
in the midplane of the galaxies.  Strong local variations of these
quantities within a single galaxy (e.g.\ near HII regions, in shocks,
etc) certainly can alter the properties of the ISM.}, and show no
signs of undergoing a significant change at $V_c=120\kms$.  We do not
discuss these parameters further here, but include them in an Appendix
for completeness.  We find several other quantities which do correlate
well with rotation speed (including stellar surface density
[\S\ref{surfdenssec}], stellar velocity dispersion
[\S\ref{veldispsec}], shear [\S\ref{kappasec}], and metallicity
[\S\ref{metallicitysec}]), but none of these showed a sharp,
discontinuous change at $V_c=120\kms$ either.  Of all the parameters
we tested, the only one which showed an obvious connection to the
observed transition is disk instability.  In particular, we find that,
above a rotation speed of $V_c=120\kms$, the galaxies in our sample
become gravitationally unstable, as parameterized by Rafikov's (2001)
formalism for evaluating radial instabilities in a mixed star$+$gas
disk.

The stability of the disk is controlled by the velocity dispersion and
the self-gravity (i.e.\ surface density) of both the stellar and
gaseous components, as well as by the kinematic shear within the disk.
We therefore evaluate the stability of the galaxies in our sample by
calculating the surface mass densities of the stars and gas
(\S\ref{surfdenssec}), the velocity dispersion (\S\ref{veldispsec}),
and the epicyclic frequency (\S\ref{kappasec}), using our NIR data and
existing data from the literature.  We combine these quantities to
assess the stability of our sample galaxies in \S\ref{Qeffsec}.

\subsection{Surface Density}              \label{surfdenssec}

The surface densities of both gas and stars contribute to disk
instabilities.  While the gas component tends to dominate the
instability, the stellar component provides additional mass which can
participate in and reinforce any growing perturbation.  In this
section we calculate the mass surface density of each component.

To calculate the stellar surface density of our sample galaxies, we
parameterize the $K_s$ band light profile as a radially exponential
disk viewed in projection with an isothermal vertical velocity
dispersion: $\Sigma(R,z) = \Sigma(0,0) {\rm sech}^2(z/z_0) (R/h_r)
K_1(R/h_r)$, where $h_r$ is the radial exponential scale length, $z_0$
is the vertical scale height, and $K_1$ is the modified Bessel
function of the first kind.  We then fit for observed peak surface
brightness $\Sigma(0,0)$, the radial exponential scale length $h_r$,
and the vertical scale height $z_0$, and then solve for the face-on
central surface brightness $\Sigma_0=\Sigma(0,0) (z_0/h_r)$ (van der
Kruit \& Searle 1981).  Details of the fitting procedure are described
in Yoachim \& Dalcanton (2004).  Similar work by Barteldrees \& Dettmar (1994)
and Kregel et al.\ (2002) has shown that this procedure produces
measurements of $z_0$ and $h_r$ which are accurate to $\sim\,$10\% for
inclinations greater than $\sim\,85\degree$.  We also assume a
constant disk scale height with radius, consistent with the results of
de Grijs \& Peletier (1997) for late-type disks.  Finally, we convert
surface brightness to surface density by adopting a color-dependent
stellar mass-to-light ratio of
$\log_{10}(M/L)_{K_s}=-0.776+0.452(B-R)$ in solar units, based on the
results from Bell \& de Jong (2001).  This yields a mean of
$(M/L)_{K_s}=0.49\pm0.14\msun/\lsun$ for the sample.  While the
resulting stellar mass-to-light ratios are robust for relative
measures of the stellar mass density, the absolute value may be
subject to systematic shifts depending on differences between the
assumed and the true underlying initial mass function.

To calculate the gaseous mass surface density for our sample galaxy,
we are forced to rely on estimating the HI surface density from
single-dish observations, because the galaxies in our sample have not
yet been mapped at 21 cm.  For the $\sim$1/3 of the galaxies lacking
single-dish HI observations, we estimate the HI mass $M_{HI}$ by
scaling the stellar mass $M_{stars}$ (derived from the $K_s$ absolute
magnitude using the color-dependent mass-to-light ratio above) by the
median ratio of $M_{HI}/M_{stars}$ ($=1.97$) for the other galaxies in the
sample.  We derive an approximate HI scale length by
adopting the mean ratio of HI to optical scale lengths from Swaters et
al (2002) and scaling the value of $h_r$ derived from the 2-d surface
brightness fits.  We then derive a first order estimate of
$\Sigma_{HI}=M_{HI}/2\pi h^2_r$.  However, the HI surface density
profiles of comparable late type galaxies in Swaters et al (2002) all
have central depressions, so that the central HI surface densities we
derive from this relation are significantly higher than seen in any
face-on system.  We therefore apply a second correction based on the median
ratio of the actual central surface density to the one derived in the
manner above ($0.28\pm0.12$), again using the face-on late type
galaxies in Swaters et al (2002).  Finally, we clip any values of
the HI surface density greater than $15\msun/\pc^2$, which is an
empirical upper limit for late-type galaxies.  The resulting HI
surface densities are comparable to those seen in the Swaters et al.\
(2002) sample.  We have propagated errors associated with the
derivation of $\Sigma_{HI}$, and while they are large individually,
they are small enough that the data reproduces the identical
relationship between HI surface density and stellar surface brightness
found by Swaters et al.\ (2002; see their Figure 11).

We also include a correction for the surface density of molecular gas,
which makes an important contribution to the gaseous surface density
of high mass galaxies.  There are currently no CO observations of our
sample galaxies, so we are forced to make a purely statistical
estimate of the molecular gas content based on the rotation speed of
the galaxy.  For this we have used the central surface density in
H$_2$ of nearby galaxies, as a function of their inclination-corrected
rotation speed ($V_c\equiv W_{50}/2$, as for the galaxies in our
sample) from Rownd \& Young (1999).  To approximate the selection
criteria of our sample, we have eliminated galaxies that (1) are type
Sab or earlier, (2) have peculiar or spindle morphologies, (3) are
identified as mergers, or (4) are located in close pairs.  We fit
these data with a power law relationship between $\Sigma_{{\rm H}_2}$
and $V_c$, including a Baysean treatment of upper limits.  If Virgo
galaxies are included, $\Sigma_{{\rm H}_2}=(V_c/47.1\kms)^{2.49}$ in
$\msun/\pc^2$.  When Virgo galaxies are excluded, the relation shifts
to slightly higher values: $\Sigma_{{\rm H}_2}=(V_c/22.7\kms)^{1.77}$
in $\msun/\pc^2$.  There is scatter of about a factor of 2 around this
relation at the high mass end.  The scatter increases to a factor of 5
at lower masses due to a large number of CO upper limits among the low
mass galaxies.  While the scatter is large for these low mass
galaxies, we find that the contribution from H$_2$ to the gaseous
surface density is negligible, modulo uncertainties in the CO-to-H$_2$
conversion factor.  Thus the uncertainty in the H$_2$ surface density
makes little contribution to the error budget near the $V_c=120\kms$
transition in which we are interested.

\begin{figure}[t]
\includegraphics[width=3.5in]{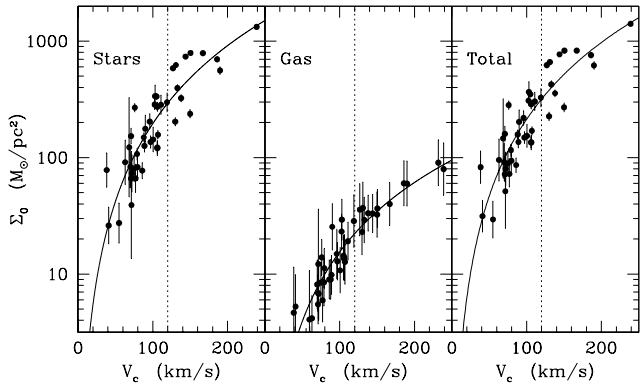}
\caption{\footnotesize 
  Face-on central stellar surface mass density (left), gas mass
  density (center), and total (stars+gas) mass density (right) as a
  function of rotation speed for the galaxies in the Dalcanton \&
  Bernstein (2000) sample.  The gas mass density includes both neutral
  and molecular components, corrected for metals.  The solid lines
  show the power law relations $\Sigma_{stars}=(V_c/9.3\kms)^{2.23}$,
  $\Sigma_{gas}=(V_c/25.1\kms)^{1.98}$, and
  $\Sigma_{tot}=(V_c/8.7\kms)^{2.19}$ in units of $\msun/\pc^2$.
  \label{surfdensfig}}
\end{figure}

The resulting central stellar mass surface density, gaseous mass
surface density ($\Sigma_{gas}=1.33(\Sigma_{HI}+\Sigma_{{\rm H}_2})$,
using the Virgo data calibration for $\Sigma_{{\rm H}_2}$ and
correcting for helium), and total baryonic surface density
($\Sigma_{tot}=\Sigma_{stars}+\Sigma_{gas}$) are shown in
Figure~\ref{surfdensfig}, as a function of rotation speed.  All
surface densities show strong correlations with rotation speed
(Spearman correlation coefficients of $>0.87$):
$\Sigma_{stars}=(V_c/9.3\kms)^{2.23}$;
$\Sigma_{gas}=(V_c/25.1\kms)^{1.98}$; and
$\Sigma_{tot}=(V_c/8.7\kms)^{2.19}$ in units of $\msun/\pc^2$.
However, none of these show a sharp change at $V_c=120\kms$.  We
note that there are many galaxies on either side of the $V_c=120\kms$
transition with comparable gas and stellar surface densities,
suggesting that a drop in surface density alone is not responsible for
the significantly increased thickness of the dust layer in low mass
galaxies.  In addition, the correlation with stellar surface density
is comparable to that seen in recent analyses of SDSS data (Kauffmann
et al. 2003, Blanton et al.\ 2003), assuming a power-law relationship
for the baryonic Tully-Fisher relation (Bell \& de Jong 2000).

\placefigure{surfdensfig}

Finally, we approximate the dependence of surface density on radius by
assuming that the surface density of both the gas and the stars fall
of radially as $e^{-r/h_r}$ (i.e.\ as an exponential disk; Regan et
al.\ 2001).

\subsection{Velocity Dispersions}                       \label{veldispsec}

\begin{figure}[t]
\includegraphics[width=3.5in]{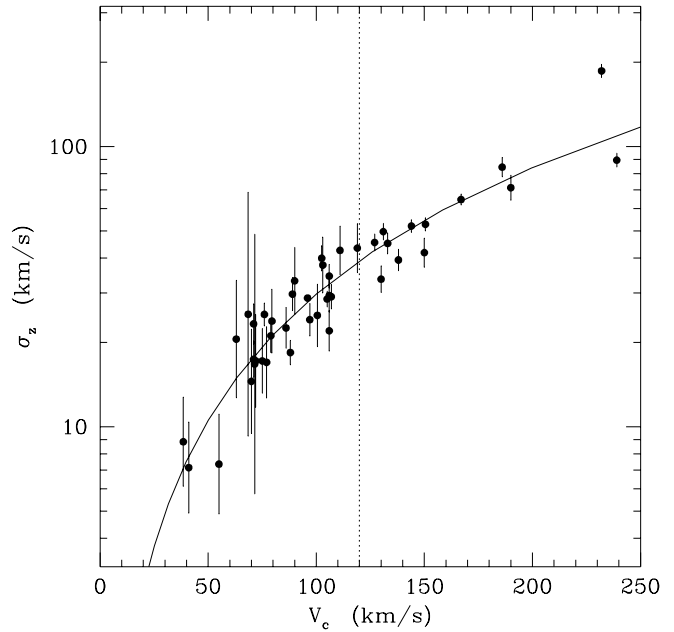}
\caption{\footnotesize 
  Vertical velocity dispersion derived from 2-d fits to the
  $K_s$ band surface brightness distribution, as a function of
  rotation speed, for the Dalcanton \& Bernstein (2000) sample.  The
  dashed line shows the relation $\sigma_z = (V_c/10.4\kms)^{1.5}$.
  \label{veldispfig}}
\end{figure}

We derive the vertical velocity dispersion of the stars by measuring
the scale height which results from the balance between dynamical
pressure and the gravitational force of the disk.  For the isothermal
disk population used in the 2-d fits the vertical velocity dispersion
is $\sigma_z^2= \pi G \Sigma_{stars} z_0$, neglecting the surface
density of the gas.  We include an approximate correction for the
effect of the gas by using $\Sigma_{tot}$ instead of $\Sigma_{stars}$
in the expression for $\sigma_z$; although this is not exact, the gas
density is negligible for nearly all of our sample and the
approximation is adequate for this calculation.  The resulting
vertical stellar velocity dispersions are shown in
Figure~\ref{veldispfig}. There is an extremely tight power-law
correlation between velocity dispersion and rotation speed, $\sigma_z
= (V_c/10.4\kms)^{1.5}$ in $\kms$, with a Spearman correlation
coefficient of 0.92.  However, no sharp transition is seen at
$V_c=120\kms$.

\placefigure{veldispfig}

Because Rafikov's (2001) analysis quantifies the stability of the disk
to radial perturbations, the vertical component of the velocity
dispersion is less critical than the radial.  We estimate the radial
velocity dispersion by assuming $\sigma_r=2\sigma_z$, i.e.\ that the
3-dimensional velocity dispersion is anisotropic, which holds for the
old stellar populations of the Milky Way.  Although there are no
direct measurements of this ratio in other galaxies, there are
indirect arguments by van der Kruit \& de Grijs (1999) and van der
Kruit et al.\ (2001) which suggest comparable ratios hold outside the
Milky Way as well.  Finally, we adopt a radial dependence for the
stellar velocity dispersion of $e^{-r/2h_r}$, as observed by Bottema
(1993).

For the gas, we adopt a fixed velocity dispersion of $10\kms$ for the
HI, typical of the line width of the warm HI component in late type
and dwarf galaxies (Young \& Lo 1997, Braun 1997, Petric \& Rupen
2001, Hoffman et al.\ 2001), assuming that the warm component
dominates the neutral medium and that the gas velocity dispersion is
isotropic. We also adopt a velocity dispersion of $5\kms$ for the
H$_2$ (Malhotra 1994, Combes \& Becquaert 1997).  We then calculate
the effective gas velocity dispersion as the quadrature sum of the
velocity dispersion of each component, weighted by their relative
surface densities.  This allows us to treat the gas as a single
component in the stability analysis.  Observations of face-on disks
suggest that the velocity dispersion of HI and H$_2$ varies weakly
with radius, and thus we assume a constant velocity dispersion
throughout the disk.

\subsection{Epicyclic Frequency}                        \label{kappasec}

The epicyclic frequency $\kappa$ 
is a measure of the Coriolis force within a rotating disk.  The
Coriolis force in a strongly shearing disk can disrupt growing
perturbations and stabilize the disk for large values of $\kappa$.  We
now calculate $\kappa$ from dynamical measurements of late-type
galaxies.

Ideally, we would use spatially resolved velocity maps to measure
$\kappa$ directly for the galaxies in our sample.  Although we have
H$\alpha$ rotation curves for $\sim$2/3 of our sample (Dalcanton \&
Bernstein 2000b), those curves cannot be used due to projection
effects, the sparse sampling of HII regions, and extinction.  We
therefore derived $\kappa$ as a function of rotation speed for an
extensive sample of H$\alpha$ rotation curves of less inclined
galaxies, for which far more accurate values can be derived than from
our edge-on sample.

To measure $\kappa$ as a function of radius we calculate $V_c(r)$ and
$dV_c(r)/dr$ from the parameterized rotation curves from Courteau
(1997) (see Courteau's eqn 2), after restricting the sample to only
unbarred galaxies of type Sbc or later that have high-quality
parameterized fits to the rotation curves.  We then calculate
$\kappa(r)$ using $\kappa^2(r) = (V_c(r)/r)^2[(r/V^2_c(r))dV^2_c(r)/dr
+ 2]$ for the resulting sample of 181 galaxies.  We find that
$\kappa(r)$ varies systematically with rotation speed, with more
slowly rotating galaxies having smaller values of $\kappa$.  For
reference, we find that at one disk scale length ($r=h_r$),
$\kappa(h_r)=(0.30 \pm 0.07 \kms/\kpc)(V_c/\kms)$, where $h_r$ is
adopted from Courteau (1996).  This trend reflects that the internal
dynamics of low mass galaxies more closely approximate solid-body
rotation.

We derive the radial variation of $\kappa(r)$ by averaging together
$\kappa(r)/(V_c/h_r)$ for all Courteau (1997) galaxies that have
rotation speeds within $\pm12.5\kms$ of the rotation speed of each
galaxy in our sample.  We then scale each resulting average
curve for $\kappa(r)$ by the value of $V_c/h_r$ for the corresponding
galaxy from our sample.

\subsection{The Radial Variation of Disk Stability}          \label{Qeffsec}

We can combine the results of \S\S\ref{surfdenssec}-\ref{kappasec} to
evaluate the stability of the disks as a function of radius.  For this
we adopt the formalism of Rafikov (2001), who evaluates the stability
of a thin two component disk where the gas behaves as a cold fluid,
and the stars are dynamically warm and collisionless.  Rafikov (2001)
calculates the locus separating regions of stability (bottom) and
instability (top) on a plane of $1/Q_{gas}$ vs $1/Q_{stars}$, where
$Q_i=\kappa\sigma_i/\pi G
\Sigma_i$ for the gaseous and stellar velocity dispersions
$\sigma_{gas},\sigma_{stars}$ and mass surface densities
$\Sigma_{gas},\Sigma_{stars}$, respectively.  The exact position of the
locus depends on the ratio of gas to stellar surface density,
$R=\Sigma_{gas}/\Sigma_{stars}$, and can be solved for numerically.

\begin{figure}[t]
\includegraphics[width=3.5in]{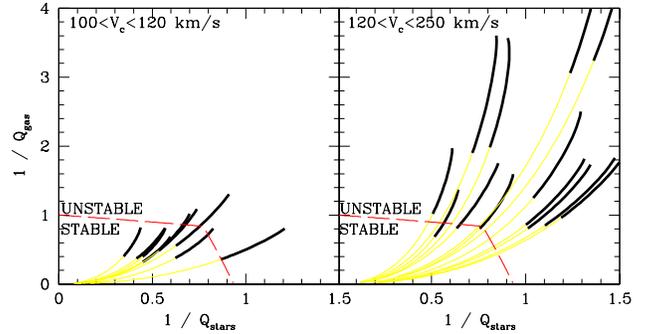}
\caption{\footnotesize 
  Disk stability as a function of radius, in two bins of galaxy
  rotation speed $V_c$.  Each line traces the radial variation of
  Rafikov's (2001) $1/Q_{gas}$ and $1/Q_{stars}$ parameter within each
  galaxy from the Dalcanton \& Bernstein (2000) sample.  The heavy
  line indicates radii within one radial disk scale length ($r<h_r$).
  Regions of the galaxies which lie above the long dashed lines are
  unstable to radial disk pertubations.  In general the inner regions
  of massive galaxies are unstable while low mass galaxies
  ($V_c<120\kms$) are stable everywhere.  \label{Qfig}}
\end{figure}

\placefigure{Qfig}

In Figure~\ref{Qfig} we plot the radial variation of $1/Q_{gas}$ and
$1/Q_{stars}$ for the galaxies with rotation speeds above and below
$V_c=120\kms$, as in Figure~\ref{dustfig}.  We have included the locus
separating stability from instability for the median value of $R$ for
the galaxies in each subsample.  The central regions of the galaxies
($r<h_r$; bold line) indicate the region used to identify the presence
of dust lanes in Figure~\ref{dustfig}.  Note that these inner regions
are typically less stable than the outer regions.

Figure~\ref{Qfig} shows a sharp difference between the stability of
high and low mass disks in their inner regions.  The high mass
galaxies (right panel) are unstable over the entire central region
within which we can easily detect dust lanes.  In contrast, the low
mass galaxies (left panel) are almost entirely stable, all the way to
their centers.  In the few low mass galaxies that are not entirely
stable, the unstable region contains a negligible fraction of the
galaxies' light\footnote{The only exception is FGC 979,
which is unusual in many other respects. It seems to be undergoing 
a modest starburst, based on its unusually blue color
and high surface brightness.  These two facts are likely to be related
(\S\ref{SFsec}).} ($r<<h_r/2$).  While there may be small unstable
regions corresponding to local gas overdensities, Figure~\ref{Qfig}
suggests that such regions cannot be widespread within the low mass
disks.

While one may not have expected the difference in stability to be so
sharply defined, it is not surprising that the stability is a strong
function of $V_c$.  From the scaling relations derived for
surface density, velocity dispersion, and epicyclic frequency,
we expect
$1/Q_{gas} = 0.44 \times (V_c/120\kms)^{0.98}$, assuming $\sigma_{gas}=7\kms$, 
at one disk scale length.
For the stars,
$1/Q_{stars} = 0.85 \times (120\kms/V_c)^{0.27}$, which depends only weakly
on rotation speed.  However, as the stability locus shows, our sample
is in a regime where the stability of the disk depends more
strongly on $Q_{gas}$.  This suggests that the stability of late type disks at
$r=h_r$ is nearly inversely proportional to rotation speed.
The exact combination of surface densities and sizes
of our sample galaxies conspires to make this trend
particularly sharp at $V_c=120\kms$.

Our conclusion that the high mass disks are substantially and
systematically less stable than the low mass disks is unchanged, or
perhaps even strengthened, if the finite thickness of disks is taken
into account.  Calculations by Elmegreen (1995) suggest that including
finite thickness effects increases the stability most for low surface
density galaxies with high gas mass fractions, such as those in the
left hand panel of Figure~\ref{Qfig}.  Thus, thickness corrections
will tend to exaggerate the offset between high and low mass galaxy
stability seen in Figure~\ref{Qfig}.

In contrast, the inferred stabilities may be affected directly by
uncertainties in $\Sigma_{gas}$ and/or $\kappa(r)$.  While we do not
have spatially resolved HI observations from which to derive these, we
nonetheless expect the estimated values to be relatively accurate, for
two reasons.  First, the central HI surface densities of late type
galaxies show a very small absolute range ($4-10\msun/\pc^2$ at the
very extremes), with a well-defined correlation with stellar surface
density.  Our procedure for estimating surface densities from
single-dish measurements reproduces the same range and correlation
seen in comparable face-on galaxies, suggesting that while our
estimates for an individual galaxy may be uncertain, the mean is
fairly accurate.  Given that the observed range of central surface
densities varies by only a factor of two, then our uncertainty in
$\Sigma_{HI}$ must be substantially smaller.  There are additional
uncertainties in the gas mass surface density from the molecular gas
component.  The intrinsic scatter in the H$_2$ surface density is much
larger than the HI (spanning a factor of 10 at the extremes), and we
do not yet have measurements of CO fluxes.  However, molecular gas is
not the dominant gas phase for galaxies near the $V_c\sim120\kms$
transition, limiting the degree to which the larger uncertainty in the
molecular gas surface density propagates into $\Sigma_{gas}$.  In the
end, the scatter in $\Sigma_{gas}$ as a function of rotation speed is
comparable to that found for $\Sigma_{stars}$
(Figure~\ref{surfdensfig}), for which we have reasonably accurate
measurements, suggesting that our uncertainties are not so large as to
have swamped the underlying trend.  Second, more accurately measured
values of $\kappa(r)$, while desirable, are unlikely to deviate much
from the values adopted from the Courteau (1997) dataset.  At one disk
scale length, the intrinsic dispersion in $\kappa(r)/V_c$ is less than
25\%.  The scatter is somewhat larger at smaller radii, but this is
due to the wide variety of bulge strengths in the Courteau (1997)
sample -- a feature which would not affect our sample of bulgeless
galaxies.  Moreover, there is a large degree of self-similarity in the
rotation curves of galaxies of similar mass (Persic, Salucci, \& Stel
1996), and thus there are unlikely to be major errors introduced by
our adopting values for $\kappa(r)$ from a separate sample of
galaxies.

In conclusion, we find a substantial difference in the stability of
galaxies above and below $V_c=120\kms$.  When combined with the
evidence in \S\ref{datasec}, this suggests that the onset of
instabilities is likely to play a key role in the formation of dust
lanes.  

\section{How Disk Stability Shapes the Morphology of the
Cold ISM}       \label{theorysec}

We now turn our attention to the mechanism by which disk instabilities
might lead to the formation of a thin dustlane.  Key to our
interpretation is the belief that the transition in dust morphology
must involve the entire cold ISM, not just the dust component.  Thus,
the onset of dust lanes in high mass galaxies discussed in
\S\ref{datasec} must reflect a significant decrease in the scale
height of the cold ISM.  In the Milky Way, and presumably within our
sample as well, the scale height of the cold ISM is set by the
equilibrium between turbulent velocities and surface density.
Therefore, the onset of disk instabilities may potentially decrease
the scale height by altering the turbulent velocities of the gas
layer.  We now discuss these ideas in more detail.

\subsection{The Connection Between Dust and the Cold ISM}  \label{dustgassec}

The transition in the morphology of the dusty ISM is likely to reflect
a structural shift in the entire cool gaseous ISM, for several
reasons.  First, if dust is evenly mixed with gas throughout the ISM,
then the highest densities of dust will coincide with the highest gas
densities, i.e.\ in the coolest phases of the ISM.  This assumption is
based both on observational data (e.g.\ Allen, Atherton, \& Tilanus
1986, Vogel, Kulkarni, \& Scoville 1988, Rand 1995) and theory, which
suggests that dust and gas have slow relative drift velocities
(Weingartner \& Draine 2001).  The close physical coupling between
dust and gas therefore suggests that the morphology of the dust
extinction gives a first order indication of the cold gas
distribution.  Second, dust is the catalyst for the formation of
molecular gas, and thus the cold gaseous ISM will track regions of
high dust density -- i.e.\ it will be easiest for molecular gas to
form where the dust density itself is highest.  Finally, dust is
destroyed principally by sputtering and grain-grain collision in
supernovae remnants (Dwek 1998, Draine \& Salpeter 1979), and thus it
may avoid all but the cooler phases of the ISM.  These arguments all
suggest that regions with the highest levels of dust extinction will
also be associated with the cold molecular phase or its immediate
precursor, the cool neutral medium (as can be clearly seen in Figure 6
of Dame et al.\ 2001).  We therefore assume that the change in dust
morphology with disk stability implies an accompanying change in the
cool gas distribution.

\subsection{The Importance of Turbulent Velocities}           \label{turbsec}

The ground-based and HST data presented in \S\ref{datasec}, and the
dust-cold ISM connection drawn in \S\ref{dustgassec}, suggest that the scale
height of the cold ISM is significantly smaller in gravitationally
unstable high mass galaxies.  What sets this scale height?  

As discussed in \S\ref{surfdenssec} for the stellar disk, the scale
height of the ISM results from the balance between gravity and
pressure.  The sharp change observed in the ISM scale height suggests
that either the self-gravity of the disks or the pressure supporting
the ISM changes discontinuously across the $V_c=120\kms$ transition.
However, the data presented in Figure~\ref{surfdensfig} indicates that
the self-gravity of the disks is likely to be similar on either side
of the transition.  Therefore, it is much more likely that the
pressure of the ISM drops significantly with the onset of
gravitational instabilities, producing thin dust lanes. Indeed, we
have solved for the vertical equilibrium structure of the gas and
stars and find that we cannot produce a sharp transition in the
relative thickness of the gas and stars without assuming that the gas
pressure is systematically lower in the unstable high mass
galaxies. The question then becomes, what is the mechanism for this
pressure support, and why might it drop when disk instabilities are
present?

Within the ISM, the dominant source of pressure is dynamical
($P\sim\sigma_{gas}\rho$), and results from random motions of the gas,
rather than thermal pressure.  These motions are thought to arise
almost entirely from turbulence, consisting of superposed ``eddies''
with a range of physical scales and velocities.  The amplitude of
these velocity fluctuations vary as a power law over several decades
in physical scale (see review by V\'azquez-Semadeni 1999).  For
classical Kolmogorov turbulence of an incompressible adiabatic fluid,
this power law spectrum develops when energy input at one particular
physical scale ``cascades'' to other scales.  One can think of this as
a process that generates eddies of a particular size (i.e.\ the
``driving scale'' of energy input).  These then break into smaller
eddies, or superimpose to form larger ones.
Although the ISM is neither incompressible nor adiabatic, the velocity
fluctuations of the interstellar medium are observed to follow a
turbulent power-law spectrum over several orders of magnitude (e.g.\
Larson 1981, and more recently Ossenkopf \& Mac Low 2002).  In such a
medium, the velocity dispersion $\sigma_{gas}$ is the {\emph{rms}}
velocity of the gas integrated over all physical scales $L$.  
The exact form of the velocity spectrum $v(L)$ depends on the amount
of energy injected into the interstellar medium and the physical
scales at which that energy is injected.  This suggests that changes
in the mechanism driving the turbulence may alter the turbulent
velocity spectrum, and with it, the global {\emph{rms}} velocity of
the gas.

The connection between the the {\emph{rms}} velocity of the gas and
the driving mechanism of turbulence suggests a possible mechanism
linking the likely drop in the {\emph{rms}} velocity of galaxies with
$V_c>120\kms$ to the onset of gravitational instabilities.  We explore
this possibility in the following section.

\subsection{Constraints on Turbulent Driving Mechanisms}

There are several mechanisms which are thought to drive turbulence in
the interstellar medium (see Mac Low \& Klessen 2003 for a review).
On relatively small scales, turbulence can be driven by the energetic
input of stellar outflows, either from protostar jets or winds from
young massive Wolf-Rayet stars, or by supernova explosions.  On larger
scales, turbulence can be driven by magneto-rotational instabilities
(Sellwood \& Balbus 1999), or through gravitational
instabilities (e.g.\ Wada, Meurer, \& Norman 2002, although
see S\'anchez-Salcedo 2001).  Although all of these energy sources may
be operating to some degree in galaxy disks, not all of them are likely
to deliver enough energy to drive the turbulence observed in the ISM.
Of these various mechanisms, only supernova shocks (Mac Low
\& Klessen 2003) and/or gravitational instabilities (Elmegreen 2003) are
thought to provide sufficient energy input.  We will only consider these
two possible driving mechanisms in the discussion which follows.

Empirically, our results strongly suggest that the characteristic
velocity dispersion of turbulence driven in the presence of
gravitational instabilities is indeed lower than that produced by
supernovae alone.  There are two possibilities that we can identify for
explaining this difference.  First is that when disk instabilities are
present, the fragmentation of spiral arms produced by the disk
instabilities becomes the dominant driving mechanism, and that this
mechanism produces turbulence with a smaller {\emph{rms}} velocity
than supernovae-driven turbulence.  The second possibility is that
supernovae are always the dominant mechanism for driving turbulence,
but that in the presence of gravitational instabilities, the ISM layer
begins to collapse and becomes denser.  At these higher ambient
densities, the supernovae feedback may become less effective at
driving turbulence, leading the ISM to reach a new equilibrium state
with smaller characteristic velocities.  We now discuss each of these
possibilities. 

\subsubsection{Turbulence Driven by Gravitational Instabilities}

In \S\ref{quantsec} we presented evidence that galaxy disks with
rotation speeds above $120\kms$ are gravitationally unstable.  What
does this imply for the structure of such galaxies and the possible
turbulent driving mechanism?  To first order, the onset of disk
instabilities is usually associated with the formation of spiral arms.
However, a theoretical analysis by Elmegreen (1991) suggests that
although Toomre's (1964) $Q$ parameter does provide some indication of
the presence of spiral arms, it is a much more reliable indicator of
the ability of such spiral arms to fragment into clumps.  Elmegreen
(2002,2003) has suggested that this fragmentation process is primarily
responsible for driving turbulence, rather than the initial formation
of spiral arms.  Observations of molecular clouds by Hartmann (2002)
lend support to this picture.  Combined with our analysis in
\S\ref{quantsec}, these works suggest a scenario where massive 
galaxies ($V_c>120\kms$) drive turbulence primarily through
fragmentation of spiral arms.

Unfortunately, we have no theoretical explanation for why the
fragmentation process should produce turbulence with a lower
{\emph{rms}} velocity dispersion than supernova-driven turbulence.  It
is not unexpected that the two processes might produce turbulence with
different {\emph{rms}} velocities, given that the former has a
characteristic velocity scale of the Jean's length divided by the
free-fall time of a typical fragment, while the latter has a velocity
scale characteristic of expanding shock waves in the ISM.  However,
the {\emph{rms}} velocity dispersion depends on the fully
developed turbulent velocity field, and we cannot give a complete
explanation for why the velocities associated with the fragmentation
process should be smaller overall.  Neither can numerical simulations
currently shed much light on this issue, due to a lack of the
computational power necessary to model 3-dimensional supersonic
dissipative turbulence on galactic scales.  Instead, we are left with
the empirical fact that, if fragmentation of spiral arms is indeed the
dominant mechanism for driving turbulence, then the resulting
turbulence appears to have lower characteristic velocities than that
driven by supernovae.  This result may guide future theoretical
efforts to understand turbulence in the ISM.

One possible question that arises in the above scenario is why
gravitationally unstable disks, which have both fragmentation and
supernovae available as turbulent driving mechanisms, would have
{\emph{lower}} turbulent velocities than stable disks, in which
turbulence is powered only by supernovae.  Instead, one might expect
the turbulence velocities to be larger when both driving mechanisms
can be operating simultaneously.

The resolution to this question is more subtle than it first appears.
Our empirical result that gravitational instabilities lead to
turbulence with lower characteristic velocities is based upon the
observed thinness of the dust layer.  The apparent scale height of
this layer depends upon the scale height of the cold gas integrated
along the line of sight and weighted by the local extinction.
Therefore, regions of the galaxy with very high extinction and/or
reddening dominate the apparent thickness of the dust lane, while
regions of the galaxy which have little cold gas produce little
extinction, and thus have no significant effect on the morphological
appearance of the dust.  The highest densities of molecular gas are
always found in spiral arms (e.g.\ Helfer et al.\ 2003), and thus
these regions will dominate the apparent distribution of dust when
viewed in the edge-on orientation.  This implies that the low velocity
dispersion we have inferred only applies to the cold gas
localized within spiral arms.

Outside of the spiral arms, the turbulence may well have a different
velocity spectrum.  Turbulence decays rapidly, and it is unlikely that
low velocity turbulence driven within a spiral arm persists outside
the arm.  After a spiral density wave passes through a region,
fragmentation ceases to be a significant driving mechanism and
supernovae and late-stage stellar evolution must take over.  However,
these inter-arm regions have a much smaller fraction of cold gas
(Crosthwaite et al.\ 2001).  They therefore produce little extinction
and/or reddening and do not significantly influence the observed
edge-on morphology of the dust distribution.  The exceptions are the
few cases where the supernovae-driven turbulence itself produces
significant amounts of cold gas in the interfaces between turbulent
eddies. Observationally, this is probably manifested in the small
subset of galaxies known to have ``extra-planar dust'' above the
confines of the dust lanes (Howk \& Savage 1999).  These systems have
particularly high star formation rates and are the rare cases where we
see evidence of both modes of turbulent driving along a single line of
sight.  Finally, in stable low mass galaxies, we again become
sensitive to the extinction caused by molecular gas created by
supernovae-driven turbulence, because it is no longer swamped by the
much larger extinction produced globally by dense spiral arms.

\subsubsection{Turbulence Driven by Supernovae when Gravitational 
Instabilities are Present}

The above scenario suggests that low turbulent velocities result when
gravitational instabilities are the dominant mechanism driving
turbulence.  However, there is disagreement in the literature about
whether or not gravitational instabilities could ever provide more
energy input than supernovae (e.g.\ Elmegreen 2003, Mac Low \& Klessen
2003).  If not, then supernovae must always be the dominant turbulent
driving mechanism.  Supposing this to be true, the empirical
correlation we find between low turbulence velocities and the
presence of gravitational instabilities indicates that the
feedback of the supernovae into the ISM is different when
gravitational instabilities are present.  In other words, even though
supernovae always are the principal drivers of turbulence, the
resulting spectrum of turbulence is different in disks which are
gravitationally unstable.  

At first glance, this interpretation is not unreasonable.  Assuming
that our evaluation of stability is measuring the ability of spiral
arms to fragment, then large regions within the arms can undergo
local gravitational collapse.  Perturbations along the spiral arms
become more massive than their Jeans' mass, and are no longer
supported by random motions.  These regions can collapse vertically,
compressing the gas to high densities, and seeding episodes of
enhanced star formation.  At some point, the feedback from the
resulting supernovae will either stabilize the collapse or disperse
the collapsing cloud, and the gas layer will find a new equilibrium
scale height.  If the degree of compression from the gravitational
collapse cannot be completely counter-acted by the resulting increased
energy input from supernovae, then the new scale height of the gas
layer will be smaller.

At second look, however, there are a few possible problems with the
above scenario.  First, the higher densities in the collapsing gas
layer should lead to an elevated star formation rate.  If anything,
this should increase the energy input of supernovae, leading to higher
turbulent velocities, rather than lower ones.  To get around this
first objection, supernovae would need to be less efficient at driving
turbulence in the dense gas of the collapsing gas layer.  One solution
would be if there is less coupling between the supernovae and the gas
because the supernovae ``blow out'' of the thinner gas layer before
depositing much energy (Mac Low \& McCray 1988).  Another solution
would be if the velocities of the supernova shock wave itself is lower
in the denser ISM.  However, the velocity of a shock wave depends
extremely weakly on the density of the ambient medium, making this
second solution unlikely.

The second possible problem with supernova-dominant turbulence is that
the {\emph{rms}} velocity dispersion of a turbulent gas tends to be
dominated by the velocities of the largest eddies.  However, the
characteristic length scale associated with supernova driving is
relatively small, corresponding to the typical spatial separation
between supernovae.  In contrast, the driving scale of gravitational
instabilities is much larger, corresponding to the Jeans' length of
the collapsing fragments along the spiral arms. Driving by
gravitational instabilities may therefore dominate at the large length
scales which contribute most to the global velocity dispersion.  While
all of these issues are potentially resolvable, full understanding
must wait for more detailed calculations and numerical simulations.

\subsection{Velocity Dispersions in Face-On Galaxies}

We have hypothesized above that the turbulent velocity dispersion of
the cold ISM is systematically lower in spiral arms.  In principle,
this difference could be measured directly in the face-on orientation,
allowing a direct comparison between the velocity dispersions of
stable and unstable disks, and inside and outside of spiral arms.
However, such observations will be complicated by the high spatial
resolution needed, and by the very real possibility that the
CO-to-H$_2$ conversion factor varies between the two different
turbulent regimes.  We have performed a preliminary analysis of the
FCRAO galaxy survey (Young et al.\ 1995) and find that almost all
stable galaxy disks are undetected in CO.  The required velocity
dispersion measurements are therefore difficult, if not impossible, to
make in stable disks.  Even in unstable disks, the degree to which CO
emissivity traces the molecular gas is unclear.  In the presence of
star-formation, the emission from CO can be artificially enhanced by
UV-heating by young stars or by low energy cosmic rays, leading the
distribution of CO to be displaced from the dust lane (e.g.\ Rand,
Lord, \& Higdon 1999).

As an alternative, the velocity dispersion of HI can be employed as a
potential substitute for the difficult CO measurements.  HI may not be
an ideal tracer of the conditions within the molecular gas, but has
the benefit of being easily detected.  Since most HI is found in
inter-arm regions of spiral galaxies, where we suspect turbulence is
driven by the same mechanisms as in stable galaxies, the global
velocity dispersion of an entire galaxy may not change much above and
below the $V_c=120\kms$ threshold.  However, our hypothesis suggests
that high resolution spatially resolved maps of HI should reveal
systematically lower velocity dispersions in spiral arms undergoing
gravitational collapse.  Observations of the spiral galaxy NGC 1058 by
Petric \& Rubin (2003) seem to find exactly this effect, giving some
preliminary support to our hypotheses.  In low mass galaxies without
spiral arms, star formation is also confined to regions with low HI
velocity dispersion (e.g.\ NGC 2366; Hunter, Elmegreen, \& van
Woerden 2001).  This suggests that {\emph{local}} gravitational
collapse may also be associated with the low velocity dispersions we
have hypothesized are associated with more global gravitational
collapse in unstable disks.

Finally, we note that the velocity differences between
instability-driven and supernovae-driven turbulence are actually not
expected to be terribly large, and thus may have easily escaped
detection.  Although we have not yet carried out a full radiative
transfer analysis to measure the thickness of the dust layer in our
sample galaxies, we estimate that the scale height is not more than a
factor of 2 larger than the scale height found in unstable disks.  For
an isothermal distribution the scale height is proportional to the
velocity dispersion squared, and thus we expect no more than a factor
of $\sqrt{2}$ difference between the velocity dispersions measured in
the arm and inter-arm regions.  This would correspond to a small change
from $10\kms$ in the inter-arm region to $7\kms$ within the arms.

\section{Star Formation and Disk Stability: Explaining the 
Kennicutt Star Formation Threshold} \label{SFsec}

Kennicutt (1989), and more recently Martin \& Kennicutt (2001), have
argued that there is a strong empirical connection between disk
stability and star formation efficiency.  Specifically, they find that
at large radii in disk galaxies there is a precipitous drop in the
star formation rate, as measured by H$\alpha$.  They argue that the
fall-off in star formation occurs where the gas disk becomes stable,
which they identify as where the surface density of the gas falls
below a critical surface density, $\Sigma_{crit} \equiv
\alpha_Q\sigma_v\kappa/\pi G$, where $\alpha_Q$ is a numerical
constant derived from observations.  In regions of galaxies where
$\Sigma_{gas}$ is low compared to $\Sigma_{crit}$ the value of $Q$
($\propto 1/\Sigma_{gas}$) is high and the disk is stable.

The existence of the Kennicutt star formation threshold implies that
there may be an underlying link between disk instability and efficient
star formation.  Simultaneously, our results suggest a second link
between disk instability and the creation of a thin layer of cold ISM.
Taken together, these empirical results suggest a possible physical
connection between a thin layer of cold ISM and efficient star
formation.  We now consider how this connection can potentially
constrain the direct physical mechanism driving the Kennicutt star
formation threshold.

The results in this paper suggest that there are two significant
changes in the state of the cold ISM when disks become unstable.
First is that the layer of cold ISM collapses vertically.  The new
equilibrium state of the gas is therefore much denser, on average,
when disk instabilities are present.  Although the Schmidt law for
star formation in disks is usually written in terms of surface density, it
can also be expressed in terms of the mass density of the gas:

\begin{equation}
\dot{\rho}_{\rm SF} \sim \epsilon_{\rm SF}\frac{\rho_{gas}}{\tau_{ff}} \propto 
               \epsilon_{\rm SF}{\rho_{gas}}^{1.5},
\end{equation}

\noindent where $\dot{\rho}_{\rm SF}$ is the star formation rate per unit
volume, $\epsilon_{\rm SF}$ is a measure of the efficiency of star
formation, and $\rho_{gas}$ and $\tau_{ff}$ are the density and the
free-fall time of the gas, respectively.  This particular form of the
Schmidt law suggests that the star formation rate per unit volume
$\dot{\rho}_{\rm SF}$ has a strong dependence on the density of the
gas, and therefore may be sensitive to changes in the scale height of
the gas layer.

Based on the dust morphologies in the HST images, we assume that the
scale height of the gas drops by roughly a factor of two when disk
instabilities are present.  This change in scale height will tend to
increase the gas density by a factor of two as well\footnote{The
dependence of gas density on scale height will not be exact, however,
since the gas remains fully turbulent.}.  The above form for the
Schmidt law therefore suggests that the star formation rate would
increase by nearly a factor of 3 when the cold gas layer collapses.
This change would {\emph{not}} be accompanied by any change in the
surface density of the gas.  The traditional expression of the Schmidt
law in terms of surface density, $\dot{\Sigma}_{\rm SF} \propto
\Sigma^{1.5}$, would therefore appear not hold across a transition
where the scale height of the gas changes significantly.  This would
lead to an apparent jump in star formation efficiency in unstable
regions of disks where the scale height of the cold ISM has dropped.

The second significant change in the cold ISM is lower turbulent
velocities when disk instabilities are present.  These lower
{\emph{rms}} velocities must accompany the sharp drop in scale height
if the surface density has not changed significantly.  The drop in
velocity dispersion may also have an
impact on the star formation rate.
In particular, under current theories of turbulent-controlled star
formation high star formation efficiencies are associated with lower
turbulent velocities.  Thus, $\epsilon_{\rm SF}$ may be significantly
larger in unstable disks, increasing the star formation rate even more
than the factor of $\sim\!3$ expected from the increase in density
alone, further sharpening the observed Kennicutt star formation
threshold.  We now review the connection between turbulent velocities
and star formation efficiency in more detail.

\subsection{Turbulent Controlled Star Formation: Velocities and Star 
Formation Efficiency}

Over the last decade there has been an increasing focus on the role
that turbulence plays in star formation.  On the one hand, random
turbulent velocities are now thought to be critical in supporting
molecular clouds against global collapse, and indeed may be a more
important source of support than magnetic fields (Klessen, Heitsch, \&
Mac Low 2000).  Thus, turbulence may play an important role in
suppressing star formation in molecular clouds.  On the other hand,
turbulence can lead to converging flows when turbulent eddies collide.
These regions have high densities, small internal velocities, and
small Jeans masses, all of which encourage small scale {\emph{local}}
collapse and fragmentation.  Turbulence can therefore enhance the star
formation rate within molecular clouds, even while suppressing
wholesale star formation by preventing the clouds global collapse (see
discussion in Mac Low \& Klessen 2003).  This emerging picture
strongly suggests that the interplay between turbulence and
gravitational collapse regulates the star
formation rate.

Current theories of turbulence-driven star formation find that the
star formation rate is sensitive to the energy spectrum of the
turbulence.  In general, when there is more turbulence on large
scales, more material can be ``swept up'' by the eddies.  A large
fraction of the mass is therefore driven into high density sheets and
filaments which then fragment into stars.  Thus, turbulent energy on
large scales is thought to increase the star formation efficiency.
However, if the turbulent velocities become too high, then
overdensities are short-lived.  Their lifetimes become shorter than the
timescale for star formation and the very high turbulent velocities
reduce the star formation efficiency.  Simulations have
confirmed both of these trends, and demonstrate that both low Mach
numbers (i.e.\ a small ratio of the {\emph{rms}} velocity to the sound
speed) and large driving scales can lead to high star formation
efficiencies (Klessen et al.\ 2000).

Recently, however, V\'azquez-Semadeni, Ballesteros-Paredes, \& Klessen
(2003) have shown that although the star formation efficiency is
correlated with both low Mach numbers and large driving scales, it
correlates even more tightly with a single number, the ``sonic scale''.
This physical length scale is the typical eddy size below which the
turbulent velocities become subsonic.  In their simulations,
velocities are significantly supersonic on large scales
and decrease steadily to subsonic velocities on small scales.  The
sonic scale is then defined as the length scale which divides these two
regimes.  When the sonic scale is large, the turbulent velocities are
supersonic over a much smaller range of scales.  Thus, large sonic
scales correspond to smaller velocities (i.e.\ lower
Mach numbers) and/or larger driving wavelengths, both of which are
associated with inefficient star formation.  Indeed,
V\'azquez-Semadeni et al.\ (2003) find a nearly one-to-one
relationship between increases in the star formation efficiency and
in the sonic scale.

In addition to being indicative of high star formation efficiencies, a
large sonic scale is also associated with a small {\emph{rms}} gas
velocity.  When averaged over all physical scales, the turbulent
velocities in a gas will have a small {\emph{rms}} value when a
smaller fraction of the gas is supersonic, i.e.\ when the sonic scale
of the turbulence is large.  This suggests that gas with globally
small {\emph{rms}} turbulent velocities has a large sonic scale,
and will therefore host more efficient star
formation.  In \S\ref{turbsec} we gave evidence for systematically low
{\emph{rms}} velocities in gravitationally unstable disks.  In light
of the results in V\'azquez-Semadeni et al.\ (2003), this suggests
that the turbulent ISM in unstable disks has a large sonic scale, and
thus will undergo efficient star formation with larger values of
$\epsilon_{\rm SF}$.

These two effects -- the larger star formation efficiency
$\epsilon_{\rm SF}$ due to lower turbulent velocities, and the
increased density $\rho_{gas}$ due to the collapsing gas layer --
combine and lead to a highly non-linear increase in the star formation
rate when disk instabilities are present.  This provides a possible
mechanism explaining the Kennicutt density threshold for star
formation, while simultaneously explaining the onset of dust lanes.

\subsection{Other Issues}

The above discussion focusses on the large view of how star formation
thresholds may operate in disks.  However, there are several more
detailed points that we now address here.

First, we note that the above discussion proposes a mechanism only for the
truncation of the star formation efficiency in stable regions of
disks, and has no immediate implications for how the star formation
rate might vary with gas density when instabilities are present (see
review by Kennicutt 1998).  Other models which presuppose efficient
formation (e.g.\ Elmegreen 2002, Kravtstov 2003) take hold in this
regime and will set the scalings between star formation rate and gas
density.

Second, in addition to the mechanism we have proposed, there are many other
models of star formation thresholds in the literature (e.g.\ Wyse
1986, , Wang 1990, Wang \& Silk 1994, Elmegreen \& Parravano 1994,
Hunter et al.\ 1998, Pandey \& de Bruck 1999, Sellwood \& Balbus
1999).  The majority of these models invoke the growth of molecular
clouds as the critical mechanism, driven by cloud-cloud collisions in
the spiral arms, or the competition between thermal collapse and
dynamical shear.  Thus they represent a distinct class of models from
the one we have explored in this paper.  These other scenarios may
play some role in establishing the exact distribution of molecular
clouds, but do not simultaneously explain the disappearance of dust
lanes with the onset of disk stability.

Finally, we address a small potential discrepancy in the above
discussion.  The Kennicutt threshold suggests that the critical value
of $Q$ separating stability from instability is reached when
$Q_{crit}=1/\alpha_Q \approx 1.4$ (Martin \& Kennicutt 2001), whereas
our analysis of disk stability assumes instead that $Q_{crit}=1$.
However, Martin \& Kennicutt (2001) have derived a value of $\alpha_Q$
based on an assumed gas velocity dispersion of $6\kms$.  Our adopted
velocity dispersions are 25\% larger, on average.  When this
difference in corrected for, we would agree with their definition of
stability to within 15\%, and we would still catagorize the stability
of the disks in our sample the same way (i.e.\ all unstable disks
would remain unstable with the slightly revised criteria for instability).

\section{The Impact of ISM Structure on Galaxy Metallicity}  \label{metallicitysec}

Empirically, the Kennicutt star formation threshold suggests that star
formation proceeds more efficiently in the presence of disk
instabilities than in their absence.  The existence of two different
regimes for star formation, coupled to the presence or absence of disk
instabilities, will lead galaxies to have different star formation and
enrichment histories, and thus to have different metallicities.  Our
data in \S\ref{quantsec} assigns the onset of disk instabilities to
galaxies with rotation speeds above $V_c=120\kms$.  It follows that
the metallicities of galaxies above and below this threshold should
differ, due to their systematic differences in star formation
efficiency.

In the first case, where disk instabilities allow the rapid formation
of dense molecular clouds, Elmegreen (2002) suggests that star
formation is extremely efficient and goes to the maximum rate
permitted by the structure of the ISM.  In this regime, any on-going
gas infall increases the star formation rate, and the new gas is
quickly consumed.  This process leads to self-regulated star
formation that maintains a nearly constant gas fraction at the edge of
instability.  In this mode the enrichment history of the galaxy will
approximate the ``balanced infall'' case of chemical evolution,
wherein galaxies maintain a constant gas fraction by consuming any
infalling gas.  This special case can be solved analytically to show
that the metallicity of the gas will approach a constant value equal
to the nucleosynthetic yield of the stellar population (see review by
Tinsley 1980).

In the second case, where disks are stable, star formation proceeds
with low efficiency.  Galaxies in this regime maintain large gas
fractions for long periods of time, enrich slowly, and
will therefore tend to have lower metallicities on average.
The metallicities will be further reduced by on-going gas
accretion.  When the star formation rate is low, the accreted gas
increases the global gas reservoir, and dilutes the metallicity.
In this case, the metallicity will have an ``effective yield'' (i.e.\
the metallicity relative to the prediction of a closed box model) that
is suppressed compared to the unstable disks.  Detailed models by
K\"oppen \& Edmunds (1999) demonstrate that galaxies with long star
formation timescales compared to the gas accretion timescale
($t_{SFR}>t_{accretion}$) will have low metallicities, low effective
yields, and larger ratios of secondary to primarily element
abundances, all in proportion to the ratio of $t_{SFR}/t_{accretion}$.

We find strong support for this picture in the data of Garnett (2002).
Garnett (2002) has shown that the metallicity of a large sample of
late-type spiral and irregular galaxies varies by a factor of 100 over
a wide range in galaxy rotation speed.  However, the galaxy
metallicities show a sharp change at $V_c\sim 120\kms$, exactly where we have
identified the transition between the two star formation regimes.
Galaxies above the threshold have constant metallicity, regardless of
$V_c$, as expected for unstable disks with efficient star formation
that balances infall.  Galaxies below the threshold show a strong
correlation between metallicity and rotation speed and have an
``effective yield'' that is systematically suppressed with decreasing
galaxy mass (assuming the underlying nucleosynthetic yield does not
vary strongly with metallicity).

Garnett (2002) and Dekel \& Woo (2002) have argued that the drop in
metallicity and/or nucleosynthetic yield in galaxies below
$V_c\sim120\kms$ reflects the onset of supernovae-driven outflows in
low mass galaxies.  However, large outflows are hard to explain in
these low mass disks.  While low mass galaxies are thought to have
``bursty'' star formation histories, their star formation rates and
efficiencies are actually small by any absolute standards (i.e.\ they
only seem high compared to their time-averaged past star formation
rates), making it unlikely that these systems experience the large
synchronous star bursts necessary to routinely drive winds (Kennicutt
1998).  Low mass galaxies also have much higher gas mass fractions
than high mass galaxies (McGaugh \& de Blok 1997), consistent with
infall being at least as important as outflow, if not more so.  Given
that the drop in effective yield occurs at exactly the rotation speed
where disks become stable, the combination of late-time infall and
inefficient star formation due to the onset of disk stability seems to
be a more plausible explanation for the observed mass-metallicity
relationship observed in disks.  Note however, that we are not
commenting on the origins of the mass-metallicity relationship in
spheroidal systems.  Massive bulges and ellipticals are not subject to
disk instabilities, and unlike most disks they do show evidence for
the short episodes of very high star formation rates required to
drive winds.

Further support for the link between disk galaxy stability and effective
yield can be found in Garnett's data, which shows a strong correlation
between effective yield and galaxy morphology.  Galaxies classified as
spirals all have nearly constant effective yields, whereas the yields
of irregulars and low surface brightness galaxies are systematically
lower (his Figure~5).  This implies that galaxies with morphological
evidence for gravitational instability throughout their disks have
high, nearly constant values of the effective yield characteristic of
efficient star formation.  In contrast, stable disks can only
experience sporadic star formation in response to local instabilities,
leading to irregular morphologies (as in dIrr and LSBs), and
suppressed yields.

Within the class of low mass stable disks, the trend of decreasing
effective yield with decreasing rotation speed is probably a
by-product of the decreasing galaxy surface density
(Figure~\ref{surfdensfig}).  While low mass galaxies are ``stable'' in
the sense that their mean surface densities are low enough that they
are unlikely to host widespread gravitational instabilities, they are
observed to host sparse localized regions of high gas density.  These
regions rise above the threshold for instabilities over small areas
and are associated with knots of star formation (e.g.\ Hunter \&
Plummer 1996, van Zee et al.\ 1996, 1997, Hunter, Elmegreen, \& Baker
1998).  Galaxies with lower surface densities require larger (and thus
rarer) density perturbations to initiate local gravitational collapse
and temporarily drive efficient star formation.  Thus, they will have
lower time-averaged star formation rates, a larger ratio of $t_{SFR}$
to $t_{accretion}$, and a lower effective yield.  The lower density of
these systems may also lead to longer accretion timescales as well,
since the infall timescale is proportional to $\rho^{1/2}$, which
would further decrease the yield. This assumption finds strong support
in the chemical evolution models of Boissier et al.\ (2001).

Finally, we comment on another more speculative mechanism that may
affect the observed correlations between rotation speed, metallicity,
and mean stellar age.  If young disk galaxies are initially gaseous,
then they are unstable for all reasonable disk parameters.  Even very
low mass galaxies with $V_c<120\kms$ will have some initial epoch of
instability, which is coupled with high star formation efficiency.
This process shuts off when enough stars have formed to lower the gas
surface density and stabilize the disk.  We have calculated the
stability of disks as a function of the fractional gas mass surface
density ($\Sigma_{gas}/\Sigma_{tot}$), using the scaling relations
from \S\ref{quantsec}, and find that this early epoch of high star
formation efficiency terminates earlier (i.e.\ at higher gas mass
fractions) in galaxies with lower total mass.  The remaining gas
reservoir is left to be processed at lower star formation efficiency.
This mechanism could potentially explain the observations of thick
disks in Paper II by producing thick disk stars during an epoch of
instability in the young gas rich disk, perhaps through the mechanism
proposed by Kroupa (2002), while still allowing the thin disks to form
at much later times.

\section{Other Consequences} \label{observationsec}

In the sections above we have discussed three empirical facts: (1) high
mass disks with $V_c>120\kms$ are unstable to gravitational
instabilities; (2) the cold ISM is systematically thicker in galaxies
with $V_c<120\kms$; and (3) distribution of dust is thicker and more
diffuse in low mass galaxies.  In addition to their implications
for star formation thresholds (\S\ref{SFsec}), these facts have other
broad implications for galaxy evolution and morphology.

\subsection{The Formation of Bulges}    \label{bulgesec}

It has become standard lore that very low mass galaxies do not have
bulges.  Recent evidence now allows us to quantify the mass scale at
which bulges disappear from galaxies.  Kauffmann et al.\ (2002) have
used the Sloan Digital Sky Survey (York et al. 2000) to examine the
dependence of a galaxy's morphology, as measured by its concentration index, 
on the galaxy's stellar mass.  They find that the concentration
index is constant below a stellar mass of $\sim10^{9.8}\msun$,
consistent with all galaxies being effectively bulgeless below this
mass (their Figure~8).  Using the baryonic Tully-Fisher relation of
Bell \& de Jong (1999), we find that this stellar mass corresponds to
a rotation speed of $V_c\sim120\kms$, exactly the rotation speed at
which the high mass disks in our sample become unstable.

We also see the same transition within our own data.  While the sample
in Paper I was chosen to consist of apparently bulgeless disks as
viewed on the Digitized Palomar Sky Survey, subsequent deep near-IR
imaging revealed the presence of small three-dimensional bulges in all
the galaxies with $V_c>120\kms$.  None of the lower mass galaxies
showed this additional component.  Further independent support can be
found in the study of dust lane thickness by Hacke, Schielicke, \&
Schmidt (1982).  They selected nearly 40 galaxies with prominent dust
lanes in the Palomar Observatory Sky Survey plates.  All of these
galaxies turned out to be type Sc or earlier\footnote{The one
exception, NGC 4758, was tentatively classified as an irregular, but
in fact is a clear merger, based on imaging from the Nearby Field Galaxy
Survey (Jansen et al.\ 2000a).}, suggesting that selecting for the
presence of dust lanes also indirectly selects for the presence of
bulges.  

Taken together, these data strongly suggest that the existence of
bulges and dust lanes in massive late type galaxies are linked.  If
so, then it is likely that the onset of disk instabilities drives not
just the formation of dust lanes but the formation of bulges as well.
This gives support to models of secular bulge formation in late-type
galaxies (Pfenniger 1993).  These models invoke angular momentum
transport and vertical heating from resonances in bars to explain the
formation of small bulges in late-type disks.  Since transient bars
are a possible by-product of disk instabilities, bulges formed by
secular evolution are less likely to exist in stable low mass galaxies with
$V_c<120\kms$.

\subsection{The Thickness of Low Mass Galaxies}	\label{thicksec}

There is a slowly growing body of evidence that low surface brightness
dwarf galaxies are thicker than normal spiral disks.
Estimates of the intrinsic axial ratios of dwarf irregulars range from
$b/a \sim 0.3$ (Hodge \& Hitchcock 1966, van den Bergh 1988, Binggeli
\& Popescu 1995) up to $b/a \sim 0.6$ (Stavely-Smith, Davies \& Kinman
1992, Sung et al.\ 1998), all of which are significantly ``puffier'' than
typical spirals (e.g.\ Kudrya et al.\ 1994).  While the past work has
pointed to significant structural differences between the disks of classic
dwarf irregulars and normal spirals, two questions have remained
unresolved.  First, what is the mass scale at which disks ``puff up''
and second, what is the responsible mechanism?

The structural analysis of our edge-on galaxy sample
(Figure~\ref{structparamfig}) shows that there is a striking
difference in the ratio of radial to vertical scale lengths
($h_r/h_z$) above and below $V_c=120\kms$.  Galaxies above this mass
threshold have axial ratios of 5:1, with little scatter.  Low mass
galaxies, however, are much rounder on average, and show much more
scatter in their axial ratios.  Although we have not done a full shape
analysis, the observed axial ratios should be nearly representative of
the true axial ratios, given that our original sample was chosen to be
the most highly inclined galaxies in the FGC catalog.  Thus, the data
in Figure~\ref{structparamfig} suggest that galaxies with
$V_c<120\kms$ are puffier on average than more massive disks.  Given
that these galaxies are stable, they are unlikely to host well-defined
spiral arms and would probably be classified as dwarf irregulars in a
face-on orientation.  Thus, they are comparable to the thick dwarf
irregular and low surface brightness galaxies considered in previous
shape analyses.

We believe that the increase in thickness in low mass galaxies may be
closely associated with the disappearance of the dust lanes.  We have
given evidence in \S\S\ref{datasec}--\ref{theorysec} that dust lanes
are absent in low mass galaxies because the cold ISM has a
substantially larger scale height.  This implies that stars forming
from the cold ISM will have a larger scale height as well.  This
process can be seen directly in Figure~\ref{hstfig}, where luminous
young stars show up as bright white point sources against the diffuse
background.  In the high mass galaxies, these point sources are
clustered close to the midplane, whereas in the low mass galaxies, the
young stars are distributed almost uniformly with height.  The
vertical scale height of the young stellar distribution will therefore
be substantially thicker in the low mass galaxies, leading to smaller
values of $h_r/h_z$.

We note that there have been other analyses in the literature that
have not found the sharp change in $h_r/h_z$ that we identify in
Figure~\ref{structparamfig}.  The most directly comparable are the
structural analysis of de Grijs (1998) and a reanalysis of the same
data set by Kregel et al.\ (2002).  These groups find no substantial
trend in $h_r/h_z$ with galaxy rotation speed.  There are at least two
reasons why such trends may have been masked in their data.  First,
their analysis was performed on $I$-band images that are substantially
affected by dust lanes.  This will lead to artificially large values
of $h_z$ in the high mass galaxies, masking the trend
which appears in our NIR $K_s$-band data.  Although these groups take
steps to minimize the effect of dust lanes, our own experiments with
2-d fitting of edge-on disks in the optical suggests that it is
difficult not to bias the scale height to large values of $h_z$, even
when masking the dust lanes.  Second, their sample includes a much
broader range of Hubble types, complicating the interpretation of the
observed flattenings in the high mass galaxies with large bulges.
Karachentsev et al.\ (1997) have also analyzed the apparent flattenings
of edge-on galaxies, using the axial ratio of the limiting
isophote of photographic plates.  The same issues which affect the de
Grijs (1998) sample are relevant here, and again it is not necessarily
inconsistent with our data that no trend with rotation speed was
found.  Moreover, the shape of the limiting isophote is a less robust
measurement of the flattening than fitting edge-on exponential disk
models, which better traces the distribution of the bulk of the stars.

\subsection{Tully-Fisher Relation}              \label{TFsec}

Our work suggests that the disk galaxy population may have a systematically
different star formation history, physical structure, and internal
extinction at rotation speeds less than $V_c=120\kms$.  These
differences may be observable as features in the Tully-Fisher relation.
We now explore the expected and observed changes due to the above
effects.

\subsubsection{Possible Changes in Slope}

A recent analysis of Tully-Fisher residuals by Kannappen et al.\
(2002) suggest that below $V_c\sim125\kms$, kinematically undisturbed
galaxies fall systematically below the Tully-Fisher relation defined
by high mass galaxies.  This would be consistent with their having low
star formation rates due their stability, as discussed in
\S\ref{SFsec}.  

\subsubsection{Possible Changes in Scatter}

The scatter and slope of the Tully-Fisher relation depends relatively
sensitively on the exact extinction model used to correct a sample of
galaxies to a face-on luminosity (e.g.\ Pierini 1999).  Variations in
the distribution of dust therefore have an effect on the Tully-Fisher
relation.  However, upon further consideration the absence of dust
lanes in low mass galaxies may have less impact on the scatter in the
Tully-Fisher relation than expected.  In galaxies with narrow,
optically thick dust lanes, the far side of the galaxy will be
unobservable
This leads galaxies with dust lanes to have luminosities reduced by
nearly a factor of two at wavelengths where the dust lane is optically
thick.  However, this change in luminosity should be relatively
consistent from galaxy to galaxy, leading to a well-defined
inclination-dependent correction for extinction and thus little added
scatter in the Tully-Fisher relation for galaxies with $V_c>120\kms$.

In contrast, when the dust has a larger scale height and is more
diffuse overall, the overall extinction might be expected to be lower,
requiring a different inclination correction for low mass galaxies.
However, the extinction may still be significant, or indeed even
larger than for a dust lane, particularly if stars are physically
correlated with dust (Bianchi et al.\ 2000).  Variations in the
distribution of dust may therefore not necessarily lead to a
noticeable change in the slope of the Tully-Fisher relation.  There
may, however, be an increase in scatter, because the inclination
correction for a clumpy dust distribution with large scale height is
quite different than for a traditional dust lane morphology (Bianchi
et al.\ 2000, Misiriotis \& Bianchi 2002).  Observational limits on
these effects have been explored by Pierini (1999), who find no trends
in the Tully-Fisher relation which exceed the uncertainties in
extinction corrections and population incompleteness.  However, his
Figure 3 may suggest an increase in scatter at rotation speeds less
than $V_c\sim120\kms$.

In summary, it seems that the drop in star formation efficiency below
$V_c=120\kms$ may indeed be having an effect on the slope of the
Tully-Fisher relation for kinematically undisturbed, non-starburst
galaxies.  However, the change in the distribution of dust has a much
smaller effect, producing no obvious change in slope, and a small to
non-existent change in scatter at rotation speeds below $V_c=120\kms$.

\subsection{Implications for High Redshift}        \label{hizsec}

The present day $V_c=120\kms$ threshold for disk stability may not
hold at higher redshifts, due to evolution in galaxies' surface
densities, gas fractions, dark matter halos, and disk velocity
dispersions.  Evolution in all of these quantities could potentially
lead to evolution in the mass scale associated with disk
instabilities, dust lanes, and efficient star formation.  However,
numerical simulations indicate that massive disks evolve at constant
$Q$ (Mayer et al.\ 2001), suggesting that the $V_c=120\kms$ threshold
for stability may be roughly the same at all redshifts.  It may also
be possible to make {\emph{a priori}} predictions for the redshift
evolution of disk stability with mass by using semi-analytic models
for galaxy formation.  However, the associated uncertainties will be
large.

Rather than attempting to predict disk stability as a function of
redshift, it may be more fruitful to use the presence of dust lanes in
high redshift disks to place empirical constraints on the evolution of
disk stability.  One could measure the rotation speeds of high
redshift disks that host dust lanes, and use the minimum observed
rotation speed as an estimate of the scale where disk instabilities
become prevalent.  The observed evolution in this velocity scale can
then be used as a constraint on models of disk galaxy evolution, by
providing an indirect measurement of disk stability at high redshift.

\subsection{The Magellanic Clouds}   \label{magellanicsec}

We note that evidence for a transition in dust morphology has been
seen in the Magellanic clouds.  In a multicolor survey of the LMC,
Zaritsky (1999) finds that the population of cool, presumably older,
stars shows a tail toward high extinction.  He argues that the heavily
extincted stars are from the far side of the galaxy, viewed through an
intervening dust lane.  In contrast, comparable stars in the SMC show
no such high extinction tail, suggesting that there is no dust lane
blocking the view of the more distant stars (Zaritsky et al.\ 2002).
The SMC is a very low mass galaxy, based on its luminosity.  Although
it has no observed rotation along the line of sight due to its face-on
orientation, it has a central velocity dispersion of $\sigma_v=21\kms$
(Hatzidimitiou et al.\ 1997) and a Tully-Fisher based rotation speed
that clearly puts it on the low mass side of the $V_c=120\kms$
transition.  The more massive LMC has a rotational speed of
$V_c=80\kms$ using the kinematic inclination of $i=33^\degree$, or
$V_c=116\kms$, using the morphological inclination $i=22\degree$ (Kim
et al.\ 1998).  Its rotation speed is therefore close to but somewhat
less than the transition speed discussed in this paper.  However, the
LMC has a strong bar indicating that the disk is currently globally
unstable, and thus may be dynamically similar to undisturbed galaxies
above $V_c=120\kms$ transition in our sample.  
The data on the Magellanic Clouds are
therefore consistent with the behavior of dust lanes inferred for
our sample.

\subsection{Evidence for a Change in the Extinction of Face-on Galaxies}

While our observations of variations in dust lane morphology have been based
on a sample of edge-on galaxies, such variations may also detectable in
the face-on orientation.  While we cannot view dust lanes
directly in this configuration, we may potentially infer their
existence through measurements of extinction and reddening, using the
reddening sensitive ratio of H$\alpha$/H$\beta$.  In almost all
conditions in the ISM, this ratio has an intrinsic value of 2.9 (Case
B recombination).  Reddening manifests itself by suppressing
the bluer H$\beta$ line relative to the red H$\alpha$ line, leading
the ratio to increase above the Case B expectation.

\begin{figure}[t]
\includegraphics[width=3.5in]{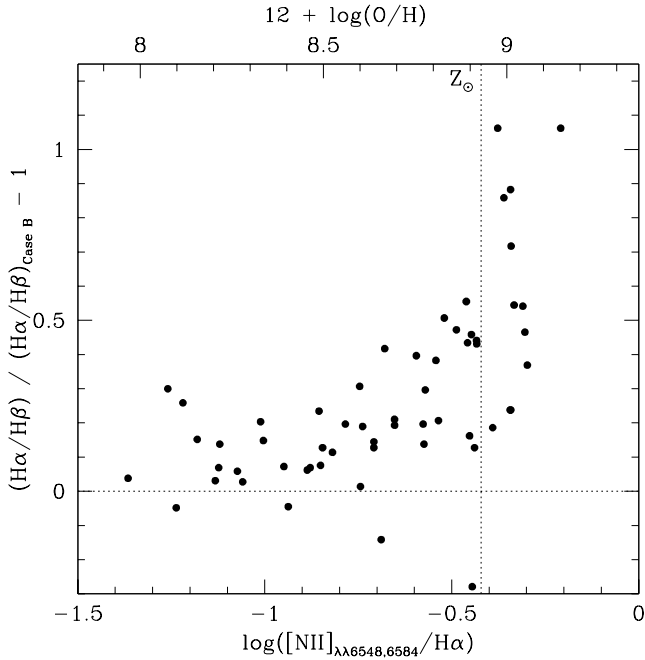}
\caption{\footnotesize 
  Reddening sensitive ratio H$\alpha$/H$\beta$ vs the metallicity
  dependent [NII]$\lambda\lambda6548,6584$/H$\alpha$ line ratio, for
  late-type galaxies from the Nearby Field Galaxy Survey of Jansen et
  al.\ (2000a,b).  Measured values of H$\alpha$/H$\beta$ have been
  scaled to the expected ``Case B'' recombination value of
  H$\alpha$/H$\beta=2.9$.  The horizonal line shows the locus of zero
  reddening; points that lie significantly above the line suffer from
  greater reddening.  The vertical line shows the solar metallicity
  gas phase abundance value (which occurs at approximately
  $V_c=120\kms$), assuming the van Zee et al.\ (1998) empirical
  conversion between [NII]$\lambda\lambda6548,6584$/H$\alpha$ and 12 +
  log(O/H); points that lie to the right of this line (higher
  metallicity) have dramatically more reddening.  \label{nfgsfig}}
\end{figure}

\placefigure{nfgsfig}

In Figure~\ref{nfgsfig}, we show the H$\alpha$/H$\beta$ ratio
(relative to Case B) for all normal late-type galaxies (Sc and beyond)
from Jansen's (2000a,b) Nearby Field Galaxy Survey.  The ratio was
derived from integrated moderate resolution spectra, and thus
represents an intensity-weighted average over the entire galaxy.  In
the absence of rotation speed information, we plot H$\alpha$/H$\beta$
as a function of metallicity using the metal-sensitive [NII]/H$\alpha$
ratio.  We use [NII]/H$\alpha$ as a proxy for rotation speed since
dynamical measurements have not yet been published for this sample.
As discussed in \S\ref{metallicitysec}, galaxies' metallicities
correlate well with rotation speed, and tend to become super-solar above
$V_c\sim120\kms$ (Garnett 2002).  The vertical line drawn at solar
metallicity therefore marks the approximate regime above which we expect
dust lanes to appear.

Figure~\ref{nfgsfig} suggests that the distribution of dust undergoes
a transition similar to that seen in edge-on galaxies.  At low
metallicities, corresponding to low rotation speeds, the degree of
reddening is quite low.  Most galaxies have mean values of
H$\alpha$/H$\beta$ within 15\% of the unreddened value.  With slightly
increasing metallicity, the ratio rises somewhat, indicating a modest
increase in reddening (about 30\% above the Case B value).  However at
larger metallicities, the reddening jumps sharply
Galaxies with super-solar gas phase abundances have a median
H$\alpha$/H$\beta$ that is nearly twice the unreddened value.  This
transition occurs at a value of [NII]/H$\alpha$ that is identical to
where we see the onset of dust lane morphologies in our edge-on
sample.

The sharp transition in the observed reddening of the face-on Jansen
et al.\ (2000a,b) sample is consistent with a rapid change in the
distribution of dust.  Although the dust content should increase with
increasing metallicity, the metallicity changes very little across the
transition near solar metallicity and we would not expect such a rapid
increase in its overall quantity.  Instead, radiative transfer
simulations have demonstrated the extreme sensitivity of extinction
and reddening to the overall distribution of dust and its degree of
clumpiness (Disney et al.\ 1989, Bianchi et al.\ 1996, Kuchinski et
al.\ 1998, Bianchi et al.\ 2000, Matthews \& Wood 2001, Misiriotis \&
Bianchi 2002).  Models in which the dust is distributed in a uniform
sheet suffer from far more reddening than those in which the same
quantity of dust is confined to small clumps mixed evenly with the
stars.  In the latter case, those few lines of sight that are heavily
reddened are also heavily extincted.  Thus, the majority of the light
that escapes is unreddened.  The uniform and clumpy distributions
assumed by these theoretical models are not exact analogs of a
turbulent dusty ISM.  However, the uniform thin dust layer model is not
a terrible analogy to the dust distribution seen along the line of
sight (i.e.\ vertically) in high mass disks viewed face-on,
particularly in the spiral arms which host most of the HII regions.
Likewise, the model with intermixed dust clumps is not a bad match to the more
vertically extended dust distribution in low mass galaxies.  Thus,
while the models are not perfect replicas of the dusty ISM, they are
likely sufficient to conclude that the sharp drop in reddening with
decreasing metallicity is more compatible with a change in dust
distribution than a sharp decrease in the overall quantity of dust.

\section{Conclusions}                    \label{conclusionsec}

Our analysis of the distribution of dust in a large sample of edge-on
late-type disks shows that the dust lane morphology appears only within
galaxies rotating faster than $V_c\approx120\kms$.  
Based on ground-based and archival HST images, we suggest that
dust lanes appear in high mass galaxies because the thickness of the
dust layer drops substantially, increasing its line-of-sight opacity and its
contrast against the stellar disk.  Dust is closely coupled to the
cold ISM, and thus these data indicate that the scale height of the
cold ISM is likely to have dropped as well.  The scale height is set
by the balance between the surface density of the disk and the
velocity dispersion of the gas.  However, the surface density does not
increase sharply above the $V_c=120\kms$ transition, and thus we
suggest that the smaller scale height is due to lower velocity
dispersions in high mass galaxies.

We have investigated possible origins for the observed transition at
$V_c=120\kms$.  We find that this transition point corresponds to the
rotation speed above which disks become gravitationally unstable.  We
speculate that the disk instabilities lead to a thin dust lane through
collapse and fragmentation of high density gas within spiral arms.  We
speculate on two possible effects which might reduce the {\emph{rms}}
velocity dispersion within fragmenting spiral arms and thus lead the
new equilibrium scale height for the gas to be lower.  The first
possibility is that the fragmentation process changes the dominant
driving mechanism for turbulence from supernovae to fragmentation.
The second possibility is that supernovae continue to be the dominant
mechanism driving the turbulent velocities, but that the resulting
velocities are lower in the dense collapsed gas layer.  Either of
these two scenarios would affect the velocity dispersion of only the
gas within the spiral arms.

We suggest that the change in dust lane morphology may provide a
physical explanation for the Kennicutt (1989) star formation
threshold.  Our data indicate that when disk instabilities are
present, the cold ISM collapses to a denser layer with smaller
turbulent velocities.  The increased density should increase the star
formation rate by roughly a factor of three, assuming a Schmidt law,
with no change in gas surface density.  In addition, numerical
simulations suggest that star formation is more efficient when
turbulent velocities are smaller, further increasing the star
formation rate.  The combination of these two effects can lead to
sharp increases in the star formation rate in the gravitationally
unstable regions of disks, reproducing Kennicutt star formation
threshold.

Our data suggest that galaxies with $V_c<120\kms$ will tend to lie
entirely below the Kennicutt threshold.  They will therefore have
systematically low star formation efficiencies and long star formation
timescales.  The long star formation timescales can potentially
explain why the metallicity and effective yield of galaxies decrease
systematically below $V_c\sim120\kms$, but are constant above.  For
the low mass stable disks, the star formation timescale is likely to
be sufficiently long that accretion can dilute the gas in between
significant star formation events.  In this regime, the metallicity
and the effective yield are systematically depressed, so that the
galaxies with the longest star formation timescales show the largest
deviations from theoretical closed box yields.  For unstable disks,
however, star formation is extremely efficient and always shorter than
the infall timescale.  In this regime, chemical evolution models
suggest that the metallicity and effective yield approach a constant
value.  This interpretation may explain the observed
trends between metallicity and galaxy mass without invoking mass loss
through outflows.

We give evidence that the onset of disk instabilities at
$V_c\sim120\kms$ may be associated with the onset of bulges at a
comparable velocity.  This gives support to theories of secular
evolution for bulge formation in late-type galaxies.  We also
demonstrate that there is a significant jump in the thickness of
stellar disks for galaxies with $V_c<120\kms$.  We argue that this
results naturally from the increase in the thickness of the cold ISM
in these stable galaxies.

In a subsequent paper we will explore the implication of our results for
the molecular gas content of galaxies.


\acknowledgements
\bigskip
\bigskip
\centerline{Acknowledgments}
\medskip

We gratefully acknowledges helpful conversations with Colin Norman,
Sangeeta Malhotra, Monika Kress on grain surface chemistry, Erik
Rosolowsky, Mordechai Mac-Low, and Bruce Draine.  In addition, we
thank Marco Spaans, for providing cooling curves below $T=10^4\K$,
Roman Rafikov for providing the stability loci for Figure~\ref{Qfig},
Stephane Courteau for providing his rotation curve data in digital
format.  Chris McKee, Carl Heiles, Connie Rockosi, Christopher Stubbs,
and Craig Hogan are also acknowledged for interesting and informative
discussions during the course of this work.  We are also grateful for
the sounding board provided by Vandana Desai, Maritza Tavarez, Andrew
West, and Beth Willman during the writing of this paper.  JJD
dedicates this paper to the memory of Dan Rosenthal.  The authors also
thank the anonymous referee for constructive comments which
substantially improved the paper.  Finally, we now acknowledges the
wisdom of E.\ L.\ O.\ Bakes who insisted that ``Dust is sublime''.

JJD was partially supported through NSF grants AST-990862 \& CAREER
AST-0238683, and the Alfred P.\ Sloan Foundation.  PY was partially
supported through UW's NSF ADVANCE program. This research has made use
of the NASA/IPAC Extragalactic Database (NED) which is operated by the
Jet Propulsion Laboratory, California Institute of Technology, under
contract with the National Aeronautics and Space Administration. 


\appendix\section{Appendix: Quantities Which Do Not Cause the Transition in 
Dust Lane Morphology}

For completeness, we briefly discuss two quantities which we rejected as
having a significant effect on the formation of dust lanes and the
morphology of the ISM.

\subsection{Luminosity Density}

\begin{figure*}[t]
\hbox{
\includegraphics[width=3.5in]{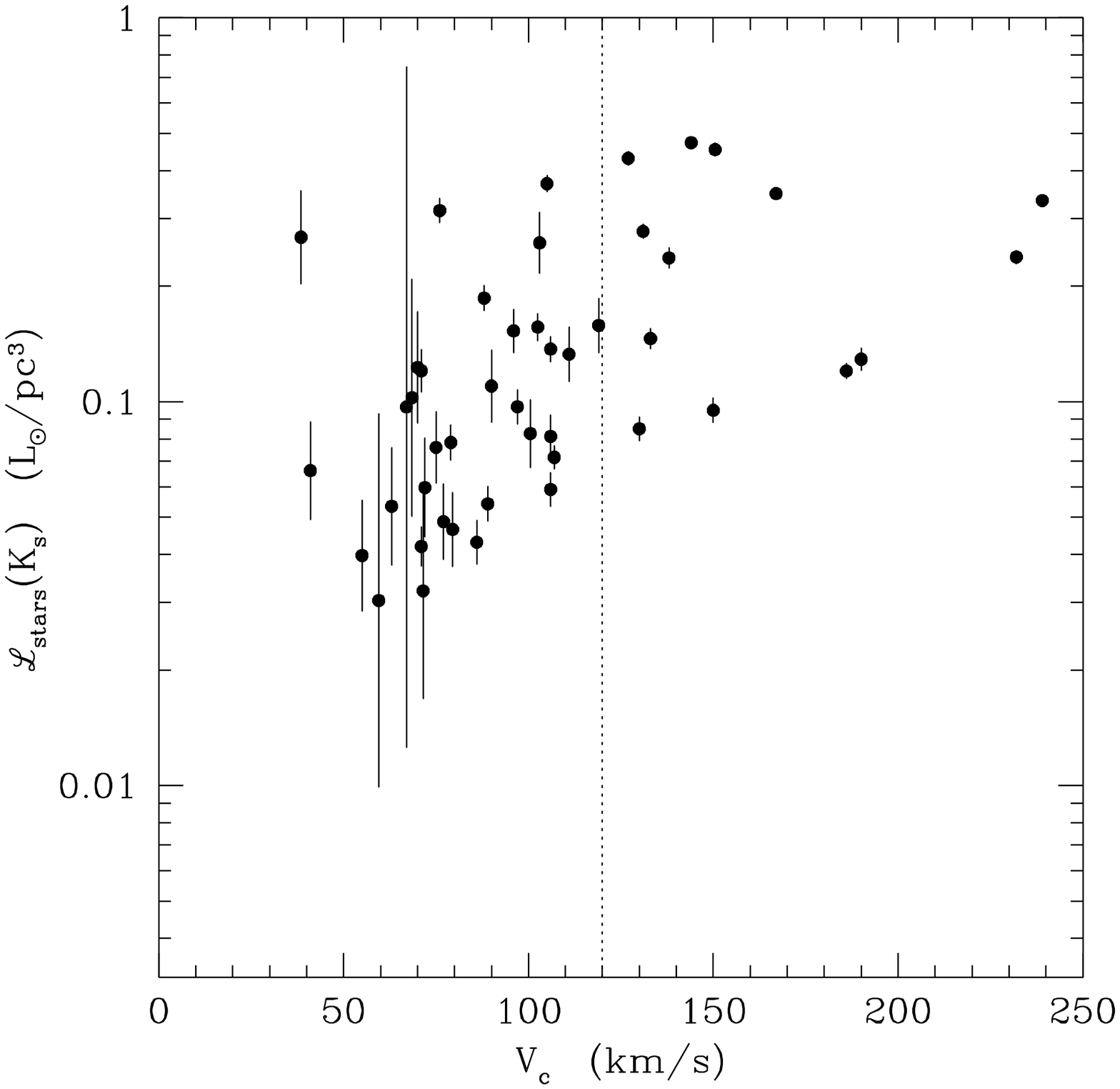}
\includegraphics[width=3.5in]{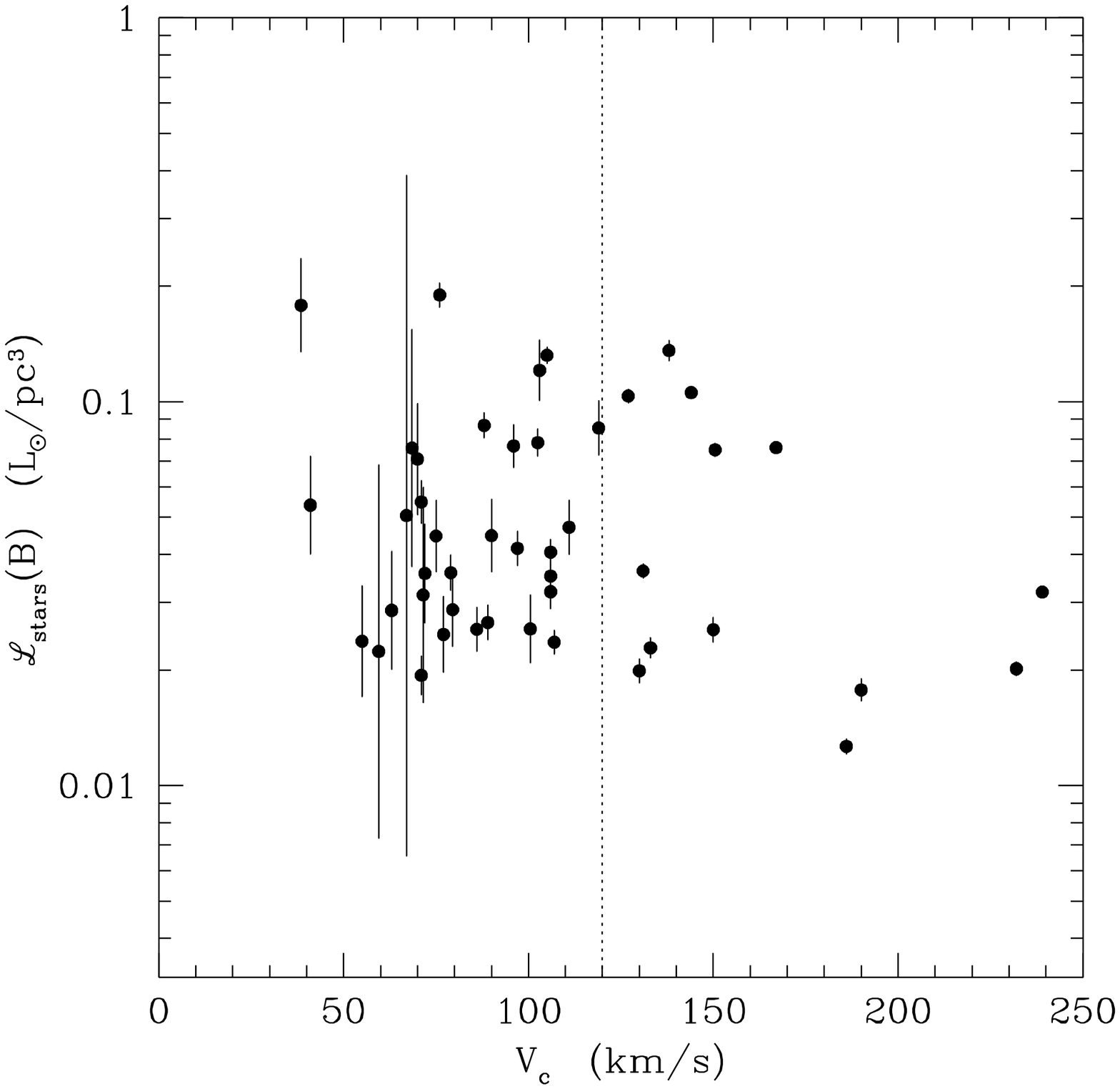}}
\caption{\footnotesize 
  Mid-plane luminosity densities in the
  $K_s$ band (left) and $B$ band (right) as a function of rotation
  speed for the galaxies in the Dalcanton \& Bernstein (2000) sample.
  There is no strong trend visible in either bandpass.
  \label{lumdensfig}}
\end{figure*}

\placefigure{lumdensfig}

In addition to being affected by surface density dependent dynamical
instabilities, the state of the ISM can be affected by the density of
ionizing radiation.  To explore this, we have used the 2-dimensional
fits to the $K_s$ band surface brightness to derive the central
midplane luminosity density, ${\cal L}_0=\Sigma(0,0)/(2h_r)$ (van der
Kruit \& Searle 1981).  The results are shown in
Figure~\ref{lumdensfig} for the $K_s$ band and for the $B$ band (the
latter giving a better estimate of the density in ionizing UV
photons).  While there is a factor of ten range in the midplane
luminosity densities, there is no statistically significant trend in
either $B$ or $K_s$.

\subsection{Internal Shear}

Another possible contributor to the observed transition is the
systematic change in the internal dynamics of the galaxies.  Shear
within a galaxy disk is a possible source of energy input into the
turbulent spectrum of a galaxy's ISM and may affect the rate at which
cold dense clouds are formed through collisions. It may also be
closely tied to the star formation rate (Hunter et al.\ 1998). If the
internal shear changes systematically as a function of galaxy mass,
then potentially the change in energy input can change the
three-dimensional structure of the ISM.

To quantify the amplitude of this effect alone, rather than its contribution
to gravitational instability, we adopt the ``universal
rotation curve'' of Persic et al.\ (1997).  Using a large database of
rotation curves, Persic et al.\ have argued that the rotation curves
of galaxies are almost entirely determined by a galaxy's rotation
speed and/or luminosity.  They have derived a fitting formula good to
roughly 10\%, allowing one to derive the approximate rotation curve
for any galaxy.  They measure the gradient of the rotation curve as a
function of a galaxy's rotation speed as well.

\begin{figure}
\includegraphics[width=3.5in]{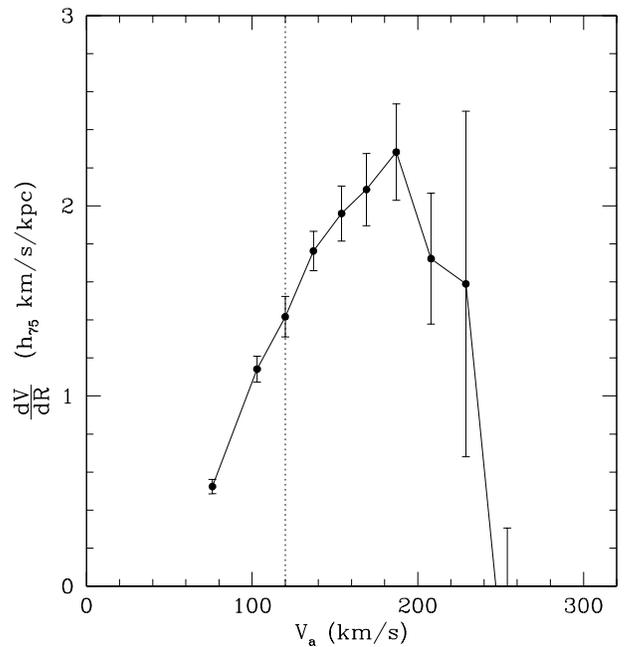}
\caption{\footnotesize The gradient in the rotation curve
  d$V$/d$R$ as a function of rotation speed $V_a$, derived from Persic
  et al's (1996) ``universal rotation curve''.  Lower mass galaxies
  have less internal shear.  The gradients are measured between
  $0.6-1R_{opt}$, where $R_{opt}=3.2h_{exp}$.  The value of $V_a$ is
  given by Matthewson et al.\ (1992).
  \label{velfig}}
\end{figure}

\placefigure{velfig}

We use the Persic et al.\ (1997) data on the gradient of the rotation
curve to derive d$V$/d$R$ in km/s/kpc, as a function of a galaxy's
rotation speed $V_a$ (from Matthewson et al.\ 1992), shown in 
Figure~\ref{velfig}.  Over the range of rotation speeds spanned by
our sample, the lower mass galaxies increasingly approach solid
body rotation, decreasing the amount of internal shear.  This suggests
that energy input due to shear is probably lower in lower mass galaxies,
decreasing the contribution that the relative internal velocities
make to the turbulent spectrum.  

The decrease in shear with decreasing mass goes in the wrong direction
for creating the observed transition, which appears to have higher
turbulent velocities in low mass disks.  We conclude that increased
energy input from dynamical shear is not directly responsible for
the disappearance of the dust lane phenomena, except for its effect on
the epicyclic frequency.  In contrast, the trend between shear and
mass does go in the right direction if creation of a dust lane
requires frequent cloud-cloud collisions, though a specific physical
mechanism has yet to be proposed.

\section{References}


\hi{Allen, R. J., Atherton, P. D., \& Tilanus, R. P. J. 1986, \nat, 319, 296}




\hi{Barteldrees, A.~\& Dettmar, R.-J.\ 1994, \aaps, 103, 475}

\hi{Bell, E. F., \& de Jong, R. S. 2000, \mnras, 313, 800}

\hi{Bianchi, S., Ferrara, A., Davies, J. I., \& Alton, P. B. 2000, 
\mnras, 311, 601}

\hi{Bianchi, S., Ferrara, A., \& Giovanardi, C.\ 1996, \apj, 465, 127}

\hi{Binggeli, B.~\& Popescu, C.~C.\ 1995, \aap, 298, 63}

\hi{Blanton, M. R. et al.\ 2003, \apj, submitted}




\hi{Bottema, R. 1993, \aap, 275, 16}


\hi{Boissier, S., Boselli, A., Prantzos, N., \& Gavazzi, G.\ 2001,
  \mnras, 321, 733}

\hi{Braun, R. 1997, \apj, 484, 637}





\hi{Combes, F.~\& Becquaert, J.-F.\ 1997, \aap, 326, 554}

\hi{Courteau, S.\ 1997, \aj, 114, 2402}

\hi{Courteau, S.\ 1996, \apjs, 103, 363}

\hi{Crosthwaite, L.~P., Turner, J.~L., Hurt, R.~L., Levine, D.~A.,
Martin, R.~N., \& Ho, P.~T.~P.\ 2001, \aj, 122, 797}

\hi{Dalcanton, J. J. \& Bernstein, R. A. 2000a, \aj, 120, 203. (Paper I)}

\hi{Dalcanton, J.~J.~\& Bernstein, R.~A.\ 2000b, ASP Conf.~Ser.~197:
Dynamics of Galaxies: from the Early Universe to the Present, 161}

\hi{Dalcanton, J. J., \& Bernstein, R. A. 2002, \aj, in press. (Paper II)}






\hi{de Grijs, R. \& Peletier R. F. 1997, \aap, 320, 21}

\hi{de Grijs, R. 1998}

\hi{Dekel, A., \& Woo, J. 2002, astro-ph/0210454}






\hi{Draine, B. T., \& Salpeter, E. E. 1979, 231, 438}


\hi{Dwek, E. 1998, \apj, 501, 643}



\hi{Elmegreen, D.~M.\ 1980, \apjs, 43, 37}


\hi{Elmegreen, B.~G.\ 1991, \apj, 378, 139}



\hi{Elmegreen, B. G. 1995, \mnras, 275, 944}


\hi{Elmegreen, B. G. 2002, \apj, 577, 206}

\hi{Elmegreen, B.~G.\ 2003, \apss, 284, 819}

\hi{Elmegreen, B. G., \& Parravano, A. 1994, \apj, 435, L121}




\hi{Fleischer, A.~J., Gauger, A., \& Sedlmayr, E.\ 1992, \aap, 266, 321}

\hi{Gail, H.-P.~\& Sedlmayr, E.\ 1988, \aap, 206, 153}

\hi{Garnett, D. R. 2002, \apj, in press}




\hi{Hacke, G., Schielicke, R., \& Schmidt, K.-H.\ 1982, Astron.\
Nachr., 303, 245}

\hi{Hartmann, L. 2002, \apj, 578, 914}

\hi{Hatzidimitriou, D., Croke, B. F., Morgan, D. H., \& Cannon, R. D.
1997, \aaps, 122, 507}


\hi{Helfer, T.~T., Thornley, M.~D., Regan, M.~W., Wong, T., Sheth, K.,
Vogel, S.~N., Blitz, L., \& Bock, D.~C.-J.\ 2003, \apjs, 145, 259}

\hi{Hodge, P. W., \& Hitchcock, J. L., \pasp, 78, 79}

\hi{Hoffman, G. L., Salpeter, E. E., \& Carle, N. J. 2001, \aj, 122, 2428}




\hi{Hunter, D.~A., Elmegreen, B.~G., \& van Woerden, H.\ 2001, \apj, 556, 773}

\hi{Hunter, D.~A., Elmegreen, B.~G., \& Baker, A.~L.\ 1998, \apj, 493, 595}

\hi{Hunter, D.~A.~\& Plummer, J.~D.\ 1996, \apj, 462, 732}



\hi{Jansen, R. A., Franx, M., Fabricant, D. 2001, \apj, 551, 825}

\hi{Jansen, R. A.,Franx, M., Fabricant, D., \& Caldwell, N. 2000a,
    \apjs, 126, 271}

\hi{Jansen, R. A., Fabricant, D., Franx, M., and Caldwell, N. 2000b,
\apjs, 126, 331}




\hi{Kannappan, S. J., Fabricant, D. G., \& Franx, M. 2002, \aj, 123, 2358}

\hi{Kauffmann, G. et al.\ 2002, astro-ph}


\hi{Karachentsev, I. D., Karachentseva, V. E., Kudrya, Y. N., Makarov,
D. I., \& Parnovsky, S. L.\ 2000, Bull. Spec. Astrophys. Obs. 50, 5}

\hi{Kennicutt, R. C. 1998, \araa, 36, 189}

\hi{Kennicutt, R.~C.\ 1989, \apj, 344, 685}

\hi{Kim, S., Staveley-Smith, L, Dopita, M., Freeman, K. C., Sault,
R. J., Kesteven, M., \& McConnell, D. 1998, \apj, 503, 674}

\hi{Klessen, R. S., Heitsch, F., \& Mac Low, M.-M.\ 2000, \apj, 535, 887}


\hi{K{\" o}ppen, J.~\& Edmunds, M.~G.\ 1999, \mnras, 306, 317}




\hi{Kravtsov, A.~V.\ 2003, \apjl, 590, L1}

\hi{Kregel, M., van der Kruit, P.~C., \& de Grijs, R.\ 2002, \mnras, 334, 646}

\hi{Kroupa, P.\ 2002, \mnras, 330, 707}

\hi{Kuchinski, L. E., Terndrup, D. M., Gordon, K. D., Witt, A. N. 1998, 
\aj 115, 1438}

\hi{Kudrya, Y.~N., Karachentsev, I.~D., Karachentseva, V.~E., \&
Parnovskii, S.~L.\ 1994, Astronomy Letters, 20, 8}


\hi{Larson, R. B. 1981, \mnras, 194, 809}

\hi{Lisenfeld, U.~\& Ferrara, A.\ 1998, \apj, 496, 145}


\hi{Mac Low, M.-M., \& McCray, R. 1988, \apj, 324, 776}

\hi{Mac Low, M.-M., \& Klessen, R. S. 2003, astro-ph/0301093}



\hi{Malhotra, S.\ 1994, \apj, 433, 687}


\hi{Martin, C. L., \& Kennicutt, R. C. Jr. 2001, \apj, 555, 301}





\hi{Matthews, L. D., \& Wood, K. 2001, \apj 548, 150}



\hi{Matthewson, D. S., Ford, V. L., Buchhorn, M. 1992, \apjs, 81, 413}

\hi{Mayer, L., Governato, F., Colpi, M., Moore, B., Quinn, T., Wadsley, J., 
Stadel, J., \& Lake, G. 2001, \apj, 559, 754}

\hi{McGaugh, S.~S.~\& de Blok, W.~J.~G.\ 1997, \apj, 481, 689}




\hi{Misiriotis, A., \& Bianchi, S. 2002, \aap, 384, 866}








\hi{Ossenkopf, V., \& Mac Low, M.-M. 2002, \aap, 390, 307}


\hi{Pandey, U.~S.~\& van de Bruck, C.\ 1999, \mnras, 306, 181}


\hi{Persic, M., Salucci, P, \& Stel, F. 1996, \mnras, 281, 27}

\hi{Petric, A.~\& Rupen, M.~P.\ 2001, ASP Conf.~Ser.~240: Gas and
Galaxy Evolution, 288 }

\hi{Pfenniger, D.  1993, in {\it Galactic Bulges,\/} IAU Symposium 153, ed.\ H. Dejonghe \& H. J. Habing, p. 387}

\hi{Pierini, D. 1999, \aap, 352, 49}



\hi{Rafikov, R.~R.\ 2001, \mnras, 323, 445 }

\hi{Rand, R. J., Lord, S. D., \& Higdon, J. L. 1999, \apj, 513, 720}

\hi{Rand, R. J. 1995, \aj, 109, 2444}

\hi{Regan, M.~W., Thornley, M.~D., Helfer, T.~T., Sheth, K., Wong, T.,
Vogel, S.~N., Blitz, L., \& Bock, D.~C.-J.\ 2001, \apj, 561, 218}





\hi{Rownd, B. K., \& Young, J. S. 1999, \aj, 118, 670}

\hi{S{\'a}nchez-Salcedo, F.~J.\ 2001, \apj, 563, 867}


\hi{Sellwood, J.~A.~\& Balbus, S.~A.\ 1999, \apj, 511, 660}


\hi{Stasi\'nska, G., \& Sodr\'e, L. Jr. 2001, \aap, 374, 919}

\hi{Staveley-Smith, L., Davies, R.~D., \& Kinman, T.~D.\ 1992, \mnras, 258, 334}

\hi{Stil, J. M., \& Israel, F. P. 1998, preprint (astro-ph/9810151)}

\hi{Stoughton, C.~et al.\ 2002, \aj, 123, 485}

\hi{Sung, E., Han, C., Ryden, B.~S., Patterson, R.~J., Chun, M., Kim,
H., Lee, W., \& Kim, D.\ 1998, \apj, 505, 199}


\hi{Swaters, R. A., van Albada, T. S., van der Hulst, J. M., \&
  Sancisi, R. 2002, \aap, in press.}





\hi{Tinsley, B.~M.\ 1980, Fundamentals of Cosmic Physics, 5, 287}


\hi{Tully, R.~B.\ 1988, ``Nearby Galaxies Catalog'', Cambridge 
and New York, Cambridge University Press}

\hi{van den Bergh, S.\ \& Pierce, M. J. 1990, \apj, 364, 444}

\hi{van den Bergh, S.\ 1988, \pasp, 100, 344}


\hi{van der Kruit, P. C., \& de Grijs, R. 1999, \aap, 352, 129}

\hi{van der Kruit, P.~C.~\& Searle, L.\ 1981, \aap, 95, 105}

\hi{van der Kruit, P.~C., Jim{\'e}nez-Vicente, J., Kregel, M., \&
Freeman, K.~C.\ 2001, \aap, 379, 374}

\hi{van Zee, L., Haynes, M.~P., Salzer, J.~J., \& Broeils, A.~H.\ 1996, \aj, 
112, 129}

\hi{van Zee, L., Haynes, M.~P., Salzer, J.~J., \& Broeils, A.~H.\ 1997, \aj, 
113, 1618}

\hi{van Zee, L., Salzer, J. J., Haynes, M. P., O'Donoghue, A. A., \& Balonek, T. J. 1998, \aj, 116, 2805}

\hi{V{\' a}zquez-Semadeni, E., Ballesteros-Paredes, J., \& Klessen,
R.~S.\ 2003, \apjl, 585, L131}

\hi{V{\' a}zquez-Semadeni, E.\ 1999, ASSL Vol.~241: Millimeter-Wave
Astronomy: Molecular Chemistry \& Physics in Space., 161}

\hi{Vogel, S. N., Kulkarnit, S. R., \& Scoville, N. Z, 1988, \nat, 344, 402}

\hi{Wada, K., Meurer, G., \& Norman, C.~A.\ 2002, \apj, 577, 197}



\hi{Wang, Z.\ 1990, \apj, 360, 529}

\hi{Wang, B.~\& Silk, J.\ 1994, \apj, 427, 759}


\hi{Weingartner, J.~C.~\& Draine, B.~T.\ 2001, \apj, 553, 581}







\hi{Xilouris, E. M., Byun, Y., Kylafis, N. D., Paleologou, E. V.,
Papamastorakis, J. 1999, \aap 344, 868}

\hi{Yoachim, P. \& Dalcanton, J. J. 2004, to be submitted to A.J.}

\hi{York, D., et al.\ 2000, \aj, 120, 1579}

\hi{Young, L. M., \& Lo, K. Y. 1997, \apj, 490, 710}

\hi{Young, J. S., Xie, S., Tacconi, L., Knezek, P., Viscuso, P.,
Tacconi-Garman, L., Scoville, N., Schneider, S., Schloerb, F. P.,
Lord, S., Lesser, A., Kenney, J., Huang, Y.-L., Devereux, N.,
Claussen, M., Case, J., Carpenter, J., Berry, M., \& Allen, L. 1995,
\apjs, 98, 219}

\hi{Zaritsky, D., Kennicutt, R. C., Jr., \& Huchra, J. P. 1994, \apj 420, 87}

\hi{Zaritsky, D., Harris, J., Thompson, I. B., Grebel, E. K., \&
Massey, P. 2002, \aj, 123, 855}

\hi{Zaritsky, D. 1999, \aj, 118, 2824}


\end{document}